\documentclass[%
 aip,
 amsmath,amssymb,
reprint, %onecolumn %%%%% Comment onecolumn to override one column format %%%%%%%
]{revtex4-2}

\usepackage{graphicx}% Include figure files
\usepackage{dcolumn}% Align table columns on decimal point
\usepackage{bm}% bold math
\usepackage[utf8]{inputenc}
\usepackage[T1]{fontenc}
\usepackage{mathptmx}
\usepackage{etoolbox}

\usepackage{hyperref} % add hypertext capabilities
\usepackage{color,soul,ulem,makecell,tabularx,float}
\usepackage{float}
\usepackage{physics}
\graphicspath{ {./} }
\usepackage{amsmath,amssymb}
\usepackage[dvipsnames]{xcolor}
\makeatletter
\def\@email#1#2{%
 \endgroup
 \patchcmd{\titleblock@produce}
  {\frontmatter@RRAPformat}
  {\frontmatter@RRAPformat{\produce@RRAP{*#1\href{mailto:#2}{#2}}}\frontmatter@RRAPformat}
  {}{}
}%
\makeatother
\begin{document}

\preprint{AIP/123-QED}

\title[]{Reversing Annealing-Induced Optical Loss in Diamond Microcavities}

% Force line breaks with \\
\author{Vinaya K. Kavatamane}
 \altaffiliation{These authors contributed equally}
\affiliation{Institute for Quantum Science and Technology, University of Calgary, Calgary, Alberta T2N 1N4, Canada}
\author{Natalia C. Carvalho}%
\altaffiliation{These authors contributed equally}
\affiliation{Institute for Quantum Science and Technology, University of Calgary, Calgary, Alberta T2N 1N4, Canada}
\author{Ahmas El-Hamamsy}
 \affiliation{Institute for Quantum Science and Technology, University of Calgary, Calgary, Alberta T2N 1N4, Canada}
 \author{Elham Zohari}
 \affiliation{Institute for Quantum Science and Technology, University of Calgary, Calgary, Alberta T2N 1N4, Canada}
 \affiliation{
Department of Physics, University of Alberta, Edmonton, Alberta T6G 2E1, Canada}
\affiliation{National Research Council Canada, Quantum and Nanotechnology Research Centre, Edmonton, Alberta T6G 2M9, Canada}

\author{Paul E. Barclay}
  \email{pbarclay@ucalgary.ca}
 
 \affiliation{Institute for Quantum Science and Technology, University of Calgary, Calgary, Alberta T2N 1N4, Canada}

\begin{abstract}

A key challenge for quantum photonic technologies based on spin qubits is the creation of optically active defects in photonic resonators. Several of the most promising defects for quantum applications are hosted in diamond, and are commonly created through ion implantation and annealing at high temperatures and high vacuum. However, the impact of annealing on photonic resonator quality factor, a critical parameter governing their coupling to defects, has not been reported. In this work, we characterize the effect of annealing at temperatures $>1200^\circ$C in high vacuum on the quality factors of diamond microdisk resonators. We investigate the optical losses associated with a non-diamond layer formed during annealing, and use Raman spectroscopy to analyze the resonator surface morphology and demonstrate that tri-acid cleaning can restore their optical quality factors. These results show the viability of creating defects in pre-fabricated diamond resonators without degrading their optical properties.
\end{abstract}

\maketitle

\section{Introduction}

Driven by the growing promise of quantum technologies \cite{ref:dowling2003qts}, chip-based photonic devices are being developed to exploit quantum mechanics in practical settings \cite{Wang-NaturePhys-2019-IntegratedQuantumTechnologies}.  Quantum networks\cite{awschalom2021development} capable of transferring information between remote nodes through coherence-preserving channels are being pursued for applications such as distributed quantum computing\cite{Monroe-PRA-2014-ModularQuantumComputer}, communication\cite{muralidharan2016optimal}, enhanced metrology\cite{Komar2014}, and sensing \cite{kim2023nanophotonic}. These applications utilize stationary matter qubits to store and process information, and photons are employed to measure and connect qubits. While nanophotonic devices play a crucial role in their implementation, of particular importance are nanoscale optical cavities, which enable coherent coupling between photons and qubits through cavity quantum electrodynamic effects \cite{Kimble2008, Janitz2020}. 

Diamond, with its wide optical transparency window, excellent thermal conductivity, high refractive index, and low magnetic nuclear spin noise, is a leading material for realizing quantum photonic technologies \cite{Janitz2020, shandilya2022diamond} thanks to its ability to host color centers with excellent spin-optical properties for implementing qubits. For example, nitrogen-vacancy centers exhibit seconds-long room-temperature spin coherence \cite{Bar-Gill2013} and enable the realization of quantum registers\cite{Bradly-PRX-2019-NV10qubit1minuteCoherence}, quantum sensors\cite{myers2014probing,Janitz2022}, and networks \cite{Hensen2015}. Group-IV color centers \cite{Bradac2019}, such as silicon-vacancies, support narrow optical transitions desirable for coherent qubit-photon interaction, and are robust to nanostructuring of their local environment \cite{Nguyen2019}. These features have enabled a wide range of quantum networking technologies \cite{pompili2021realization, Hermans2022} that are enhanced \cite{knaut2024entanglement} by integrating diamond spin qubits within photonic and phononic resonators \cite{Faraon2013,  Schroder2017, shandilya2021optomechanical, Li2014}.

The integration of color centers within diamond nanophotonic devices is often achieved through the implantation of ions of the desired impurity, including nitrogen and group IV elements, followed by high-temperature annealing under high-vacuum \cite{Pezzagna2011}. In contrast to introducing defects during the diamond growth process \cite{Chu2014NanoLett, Schroder2017}, implantation allows color centers to be created in commonly available diamond chips, and can provide control over the spatial location of color centers, for example, within photonic \cite{Sipahigil-Science-2016-SiVplatform} and phononic \cite{Wang2020APL} resonators. 
Subsequent annealing steps form impurity-vacancy complexes and repair implantation-induced damage to the crystal lattice\cite{Pezzagna2011}. Annealing processes reaching temperatures T$\sim$1200$^\circ$C while maintaining high vacuum, followed by appropriate post-annealing surface treatment, increase the spin coherence times \cite{Yamamoto2013}, enable lifetime-limited optical linewidths \cite{Chu2014NanoLett}, and improve the photostability of diamond spin defects \cite{Sangtawesin2019}. In nanostructures, color center optical transitions are affected by nearby surfaces and strain, resulting in broadening of optical transitions \cite{jantzen2016nanodiamonds} and spectral diffusion\cite{Faraon2012, Wolters2013, Ruf-NanoLett-2019-CoherentNVinMembrane}. These effects can be mitigated using inversion-symmetric group-IV color centers \cite{Evans2016}.

Despite progress in developing recipes for creating color centers, incorporating ion implantation and annealing steps into the device nanofabrication workflow is not trivial. This is largely due to trade-offs between preserving color center properties and optimizing device performance. The hardness of diamond demands aggressive etching recipes for device patterning, which can degrade color center optical and spin coherence due to surface damage and increased lattice strain \cite{shandilya2022diamond, Kasperczyk2020}. Although it is possible for the color center device integration sequence to be reversed \cite{Sipahigil-Science-2016-SiVplatform}, with resonators patterned before the implantation step, subsequent annealing steps can harm resonator performance by inducing additional losses from changes to surface morphology and other material characteristics. Despite the need to understand the impact of annealing on diamond optical resonator performance, to the best of our knowledge, studies of this effect have not been reported.

This article addresses this gap by testing the resilience of the optical properties of diamond optical cavities when they are subjected to high-temperature, high-vacuum annealing recipes commonly used to create color centers. Non-diamond layers can be created on the diamond surface from background gas or out-gassing of the annealing system at high temperatures\cite{shevchenko2024high}, which in turn can act as a source of optical loss. To assess the impact of this effect, we characterize diamond microdisk optical resonators before and after high-vacuum high-temperature annealing, and again after a chemical cleaning step commonly used for diamond surface preparation. By measuring their optical quality factors ($Q_{\text{opt}}$) after each of these stages, we observed that annealing introduced an additional optical loss channel, which can be explained by the creation of a non-diamond layer identified as amorphous carbon by surface Raman spectroscopy. We also show that $Q_\text{opt}$ can be recovered to its pre-annealing value by cleaning the diamond using a tri-acid surface treatment procedure to remove the unwanted non-diamond layer. 

\begin{figure}[!t] 
    \centering
    \includegraphics[width=0.95\linewidth]{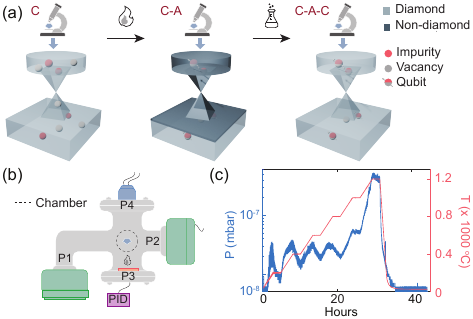}
    \caption{(a) Illustration of the protocol used to study the effect of annealing and surface treatment on diamond microdisk cavities. Optical $Q$-factors of devices were measured in pre-annealing acid-cleaned (C), annealed (C-A), and acid-cleaned stages (C-A-C). Subjecting the sample to high-temperature annealing creates a non-diamond layer represented by the darker color in (C-A). (b) Schematic of the high-temperature and high-vacuum annealing setup. P1: turbo pump, P2: ion pump, P3: heater, and P4: pressure gauge. (c) Typical time-series of sample temperature (red) and chamber pressure (blue) during the annealing procedure.}
    \label{fig1}
\end{figure}

\section{Measurement Results}

The devices studied here are fabricated from an optical grade single crystal ($\langle\text{100}\rangle$-orientation) diamond synthesized by chemical vapour deposition by Element Six. Microdisk optical cavities were patterned on the diamond chip using quasi-isotropic etching with an O$_2$-plasma to create undercut structures, a procedure first described in \cite{Khanaliloo2015}. The microdisks have diameters ranging from 5 to 8\,$\mu$m and thickness $\sim$940\,nm, and are supported by thin pedestals whose precise width increases with microdisk diameter. The optical and mechanical properties of some of the devices studied here were previously characterized in \cite{Mitchell-DiamondOptomechanicalResonator-Optica-2016}. To probe the influence of annealing and surface treatment on the microdisk's optical properties, we measured $Q_\text{opt}$ of microdisk modes at each stage of the experimental sequence shown in Fig.\ \ref{fig1} (a): cleaned (C), annealed (C-A), and cleaned post-anneal (C-A-C).  The darker color on the C-A device illustration represents the surface layer created by the annealing procedure. The schematic also shows the formation of the impurity-vacancy centers during the annealing step. These color centers are not directly studied in this work.

The annealing experiments were carried out in a home-built high vacuum annealing setup (see Fig.\ \ref{fig1} (b)) consisting of a vacuum chamber that housed a heater capable of reaching 1200$^\circ$\,C (more details about the setup in the Supplementary Material). The chamber was evacuated to pressure levels of 10$^{-9}$ mbar before heating. The temperature of the sample stage and the chamber pressure were continuously recorded throughout the annealing process, which involved gradually increasing the heater temperature from room temperature to 1200$^\circ$\,C by ramping the temperature in 200$^\circ$\,C intervals at a rate of 1$^\circ$\,C per minute, with intermediate wait times of 120 minutes between intervals to stabilize the pressure and minimize out-gassing. Finally, from 1200$^\circ$\,C the temperature was ramped down to room temperature in 120 minutes. A typical variation of heater temperature as a function of time and a corresponding variation of chamber pressure is shown in Fig.\ \ref{fig1} (c).  Prior to and after annealing, the diamond sample was cleaned using a standard tri-acid solution, which creates a highly oxidizing environment from an equal mixture of three acids: H$_2$SO$_4$, HNO$_3$, and HClO$_4$ \cite{brown2019cleaning}. The tri-acid mixture was maintained at a temperature of $\sim$250$^\text{o}$\,C for about 120 minutes (more details in the Supplementary Material). The sample was cleaned pre-annealing so that no residue on the diamond surface could be burned and thus adversely affect Q$_{\text{opt}}$.  Post-annealing, cleaning was used to remove non-diamond surface layers formed during the annealing process. Tri-acid boiling is also known to create an oxygen-terminated diamond surface through oxidation at this temperature, which is crucial for the stabilization of spin, charge, and optical properties of nitrogen-vacancy (NV) centers in diamond\cite{Janitz2022}. 

\begin{figure}[!t] 
    \centering
    \includegraphics[width=0.95\linewidth]{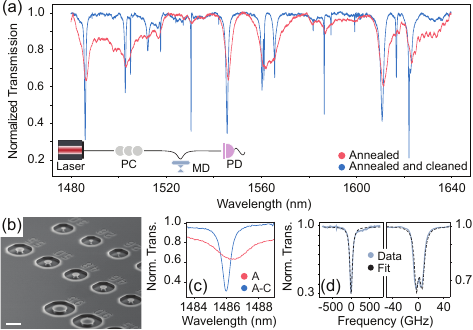}
    \caption{(a) Typical optical spectra of a measured diamond microdisk showing the optical modes of the annealed device (A) superimposed to the same modes after acid-cleaning (A-C). The left inset contains a schematic diagram of the characterization setup (PC: polarization controller, MD: microdisk and PD: photodetector). (b) Scanning electron microscope image of a typical sample (scale bar: 10 $\mu$m). (c) Zoom in on one of the optical modes shown in (a). (d) Representative data and Lorentzian fits of optical modes measured on the clean sample.}
    \label{fig2}
\end{figure}

Characterization of microdisk optical modes was performed using a swept wavelength laser (Santec TSL-710, 1480\,nm - 1640\,nm wavelength range) to measure the transmission spectra of a dimpled fiber taper evanescently coupled to individual devices. Figure \ref{fig2} (a) shows representative transmission spectra measured after annealing (A) and after annealing and cleaning (A-C), as well as a schematic of the experimental setup. Figure \ref{fig2} (b) presents a scanning electron microscope image showing some devices in a typical diamond chip fabricated with the quasi-isotropic etching technique. From transmission spectra in Fig.\ \ref{fig2} (a), we see that devices from the annealed sample support optical modes whose $Q_\text{opt}$ is significantly higher after cleaning than before cleaning. 
This is highlighted by the spectra in Fig.\ \ref{fig2} (c) of a mode of a device from the annealed sample before and after cleaning, with the latter exhibiting a narrower linewidth. Figure \ref{fig2} (d) shows representative fits of measured mode lineshapes, and shows that they can be both singlets, which are fit well with a single Lorentzian, and doublets, which are fit well by a coherent superposition of Lorentzians \cite{ref:borselli2005brs}. The doublet lineshape is characteristic of the whispering gallery modes supported by microdisks, and arises from backscattering between nominally degenerate clockwise and counter-clockwise propagating whispering gallery modes at a rate higher than their intrinsic loss rate, resulting in their renormalization into non-degenerate standing wave modes \cite{ref:borselli2005brs}. In both cases, fits to the data can be used to determine $Q_\text{opt}$ and the intrinsic quality factor, $Q_\text{opt}^i$, which removes the influence of the fiber taper on cavity linewidth, as discussed in the Supplementary Material.

\begin{figure}[!t] 
\centering\includegraphics[width=0.95\linewidth]{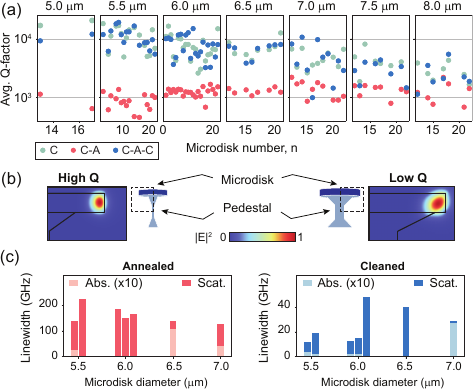}
\caption{(a) Comparison of the average intrinsic quality factors for various microdisk diameters on the same diamond chip at different stages: clean pre-annealing (C), annealed (C-A) and clean post-annealing (C-A-C). (b) Impact of the pedestal on optical losses: larger (smaller) disks have larger (smaller) pedestals and lower (higher) geometry-limited quality factors. Finite element method simulation shows the electric field of the microdisks. (c) Predictions of optical loss from absorption and scattering extracted from fitting the modified Lorentzian lineshape to resonances exhibiting photothermal effects. Each bar corresponds to measurements from a single resonance of a unique microdisk of the indicated diameter. The absorption component is multiplied by 10.}
\label{fig3}

\end{figure}

To analyze the impact of annealing and subsequent cleaning on microdisk optical loss, we measured $Q_\text{opt}$ of multiple modes from a range of devices with varying diameters when the sample is prepared in each of the three conditions described in Fig.\ \ref{fig1} (a). In Fig.\ \ref{fig3} (a) we plot the average $Q_\text{opt}^{i}$ of microdisks with diameters between 5 to 8\,$\mu$m, with each data point corresponding to $Q_\text{opt}^{i}$ averaged from several modes of the same microdisk. The influence of each process step on the average $Q_\text{opt}^i$ is clearly visible in Fig.\ \ref{fig3} (a). The cleaned microdisks before annealing (C) have modes with $10^3 \lesssim Q_{\text{opt}}^{i} \lesssim  5 \times 10^4$.  After annealing (C-A), the average $Q_\text{opt}^{i}$ of all modes was reduced to $\approx 10^3$. Finally, after the post-annealing tri-acid cleaning (C-A-C),  $Q_\text{opt}^i$  was observed to recover to pre-annealing values. These measurements demonstrate that neither the annealing nor the cleaning results in permanent damage to the optical properties of the microdisks studied here. Note that although the sensitivity of $Q_\text{opt}^{i}$ to changes in device properties varies depending on the mode of interest, the average $Q_\text{opt}^{i}$ is observed to provide a good qualitative representation of the general trend. Individual data points used to derive the average can be viewed in the Supplementary Material. 

Also evident in Fig.\ \ref{fig3} (a) is that $Q_\text{opt}^i$ for cleaned devices (C and C-A-C) trends down with increasing microdisk diameter. This is a consequence of the microdisk pedestal width's dependence on microdisk diameter due to the undercut distance decreasing with increasing device diameter in our fabrication process (See Fig.\ \ref{fig2} (b)). As a result, while the smallest diameter devices can have pedestals that are nearly fully etched away and do not interact with the microdisk whispering gallery modes, the larger diameter devices are supported by pedestals that in places are connected to the microdisk edge where the modes are confined.  This is illustrated in Fig.\ \ref{fig3} (b), which shows how the pedestal can interact with the optical field of the whispering gallery modes. Note that variations in the undercut rate along different diamond crystal plane directions affect the precise shape of the pedestal, resulting in a non-circular cross-sections that break the symmetry of the device and result in high loss if they interact with the whispering gallery mode \cite{Khanaliloo2015}.

The observed decrease of $Q_\text{opt}^i$ with increasing diameter for the cleaned devices is consistent with pedestal loss becoming the dominant loss mechanism for large microdisks: $Q_\text{opt}^i < Q_\text{ped} < \{Q_{\text{sca}}, Q_{\text{abs}}, Q_{\text{rad}}\}$, where $Q_{\text{ped}}$, $Q_{\text{sca}}$, $Q_{\text{abs}}$ and $Q_{\text{rad}}$ represent the impact of the pedestal, optical scattering, material absorption, and radiation loss, respectively.  
In contrast, in the annealed devices (C-A), $Q_\text{opt}^i$ does not vary strongly with microdisk diameter. This is consistent with absorption or surface scattering-induced loss becoming dominant, i.e. $Q_\text{opt}^i < \{Q_{\text{sca}}, Q_{\text{abs}}\} <  Q_{\text{ped}} < Q_{\text{rad}}$.

To better understand the mechanism underlying the increase in microdisk loss after annealing, it is desirable to discriminate between absorption and scattering losses for a given mode of interest\cite{ref:carmon2004dtb, Mitchell-DiamondOptomechanicalResonator-Optica-2016}. Photothermal effects arising from optical absorption within the microdisk can provide insight into the magnitude of this loss mechanism.  The small mode volume and high-$Q_\text{opt}$ of the microdisk modes can generate non-negligible local heating, and the resulting temperature rise shifts their resonance frequency \cite{ref:carmon2004dtb}. The dependence of this effect on intracavity power, which varies as the input laser is scanned across the optical resonance, creates an asymmetric lineshape (see Supplementary Material), which, when fit with a Lorentzian modified to account for this photothermal non-linearity, allows losses from optical absorption to be estimated if the device's thermal conductance is known. Figure\ \ref{fig3} (c) shows the contributions to cavity mode linewidth from absorbed and scattered optical power predicted from the lineshapes of microdisks of varying diameters using this approach. Thermal conductance values are obtained from simulations and account for variation in pedestal dimensions with microdisk diameter (see the Supplementary Material for more details). This analysis was performed on C-A and C-A-C devices, and predicts that for both before and after cleaning, surface scattering is the dominant effect limiting $Q_\text{opt}^i$.

\begin{figure}[!t] 
    \centering
    \includegraphics[width=0.95\linewidth]{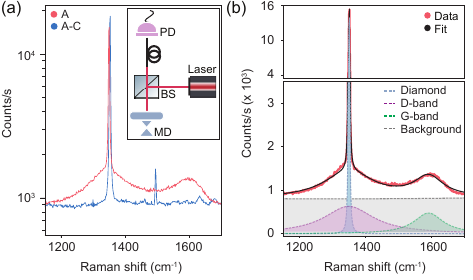}
    \caption{Raman spectroscopy of diamond surface. (a) Measurement on the freshly annealed surface (red data, A) showing a native diamond signal around 1350\,$\text{cm}^{-1}$ and a broad non-diamond peak $\sim$1600\,$\text{cm}^{-1}$. The subsequent tri-acid cleaning removes this non-diamond layer and the spectrum shows just the native diamond signal (blue data, A-C). Only the data collected for 120 seconds is shown for both cases. Inset: Schematics of the Raman spectroscopy setup. (b) Decomposition of Raman spectrum for the data A from (a) to separate different specious contributions to the signal after correcting for the background (black dotted line). The data was fit to three Lorentzian functions and the area under each peak indicates the relative contribution to the signal.}
    \label{fig:4}
\end{figure}

To further study the effect of annealing on the microdisks, we probed the optical properties of the annealed diamond material before and after cleaning using a home-built confocal Raman spectroscopy system. A schematic of this setup is presented in the inset of Fig.\ \ref{fig:4} (a). The system employs a $\sim$ 635\,nm wavelength laser for excitation, focused through a high-numerical-aperture objective (refer to the Supplementary Material for details). Both the reflected light and the Raman-scattered signals were collected via the objective and coupled into a single-mode optical fiber that functions as a confocal pinhole that rejects out-of-focus light. To maximize the Raman signal originating from the diamond surface, the focal spot was precisely aligned using the confocal microscope's depth-resolving capability. Subsequently, the Raman signal was spectrally filtered to suppress reflected excitation light before being recorded by a spectrometer. The choice of a 635\,nm excitation wavelength was deliberate to minimize interference from zero-phonon line emission of neutrally charged nitrogen-vacancy defects in diamond, which overlaps with signals from amorphous carbon phases in the Raman spectra. Data was collected across a broad spectral range (500-2500\,$\text{cm}^{-1}$) while maintaining a constant excitation laser power throughout all measurements.

In Fig. \ref{fig:4} (a), the data for both A and A-C show a sharp and distinct signal from the native diamond around 1350\,$\text{cm}^{-1}$. The spectrum obtained from the annealed surface shows the presence of a non-diamond layer in the form of a broad peak $\sim$1600\,$\text{cm}^{-1}$, whose intensity increases with signal acquisition interval. Figure \ref{fig:4} (b) shows that, as in previous studies \cite{dychalska2015study}, the Raman spectrum of A can be decomposed into contributions from D-band and G-band amorphous carbon, both of which were successfully removed through subsequent tri-acid cleaning \cite{brown2019cleaning}. Although the acid mixture does not etch diamond, the post-annealing cleaning (A-C) measurements showed a subtle blue shift ($\sim$ 0.2  nm) of the optical resonance (cavity transmission) compared to the clean sample pre-annealing (see Supplementary Material for more details). This can be explained by the amorphization of a thin layer ($\lesssim$ 1 nm) of diamond, which, when removed by the tri-acid cleaning step, reduces the microdisk diameter and thickness. As $Q_\text{opt}$ was restored during this step, we conclude that the creation and subsequent removal of this layer does not alter the microdisk surface morphology significantly in comparison to the roughness or other imperfections introduced during the initial device fabrication process.

Atomic force microscopy measurements with a  1\,nm tip radius showed RMS roughness $<$ 100\,pm in most regions, but without a marked difference between clean and annealed samples. These measurements were confined to unpatterned regions of the sample and were not able to characterize microdisk sidewall, top, or bottom surface roughness. Although mode splitting observed in clean devices (Fig. \ref{fig2} (d)) can be monitored to estimate the difference in surface roughness, the modes measured in the annealed sample do not show resolvable resonance splitting due to their broadened lineshapes.

\section{Discussion}

When integrating spin defects within nanophotonic devices, the sequence of the qubit implantation and annealing process and the device patterning process can vary depending on practical considerations. Approaches in which fabrication of devices precedes localized implantation \cite{Sipahigil-Science-2016-SiVplatform} and annealing, and approaches in which implantation-annealing precedes nanofabrication \cite{joe2024high} are both commonly employed. 
Our results show that when using the former approach, annealing of devices does not cause permanent degradation of optical resonator quality factors. This is critical for applications of high-$Q_\text{opt}$ cavities involving spin qubits, such as Purcell enhancement of photon emission and generating high optomechanical cooperativity in spin-photon-phonon interfaces \cite{shandilya2021optomechanical}.

As discussed in the experimental section, $Q_{\text{opt}}$ of our diamond microdisks is limited by surface roughness \cite{Mitchell-2019-APLphotonics-DiamondMicrodisks}. Our measurements and analysis indicate that surface roughness is increased in annealed samples as a result of the creation of an amorphous carbon layer on the device surface. Although we show that this damage can be reversed through subsequent tri-acid cleaning, it is important to note that device dimensions are modified by the removal of this non-diamond layer. This may impact cavity designs with low tolerance to geometry deviations, such as photonic crystals. We also note that if devices can be fabricated with smoother surfaces and higher $Q_{\text{opt}}$, they may become sensitive to the effect of annealing on the material's bulk optical properties and surface absorption.

Future extensions of this work can explore several open questions. One important direction is to investigate whether annealing can repair the damage induced by nanofabrication processes on photonic cavities. It is noteworthy that in devices fabricated using focused ion beam (FIB) milling, annealing in air has been reported to enhance $Q_{\text{opt}}$ through selective removal of FIB-induced damage \cite{graziosi2018single}. A closely related question is how ion implantation impacts optical losses in resonators, and to what extent post-implantation annealing can mitigate these losses. In the context of optomechanics, it is known that annealing can reduce internal friction by promoting stress relaxation in nanomechanical resonators, alleviating sources of surface \cite{wu2017reducing} and bulk \cite{metcalf2018improving} mechanical losses. Additionally, the mechanical properties of single-crystal diamond resonators are also known to be sensitive to surface termination \cite{Tao2014}. Therefore, systematic studies on how surface preparation and annealing affect nanomechanical resonances could offer valuable insights for engineering efficient spin-optomechanical platforms \cite{shandilya2022diamond}. Finally, we note that while the tri-acid cleaning used here is known to induce an oxygen-terminated surface\cite{myers2014probing,Janitz2022}, it is unclear whether this termination plays any role in the recovery of $Q_\text{opt}$ in our case. 
It would therefore be worthwhile to investigate in more detail whether diamond surface termination with different chemical species (O$_{2}$, F$_{2}$, and N$_{2}$\cite{Janitz2022}) affects $Q_\text{opt}$ of the resonators.

\section{Conclusion}

In this work, we have investigated the effect of high-temperature annealing in high-vacuum on optical losses of diamond microcavities. An analysis of sources of optical loss before and after annealing, together with Raman spectroscopy to study the properties of the diamond surface at each step, indicates that high-temperature high-vacuum annealing creates an amorphous carbon layer that adds optical loss to the cavity modes. Although this non-diamond layer reduces the cavities' optical quality factor by increasing surface roughness, our results show that it can be removed with a boiling tri-acid mixture. This demonstrates that creating color centers in pre-fabricated high-$Q$ diamond cavities is feasible without degrading device performance. This finding will aid in the creation of diamond devices that enhance coherent interactions between quantum systems required for applications such as quantum transduction and quantum networking.

\begin{acknowledgments}
We acknowledge the assistance of Joseph Losby, XueHai Tan and Peng Li at the University of Alberta \textit{nanoFAB} for XRD and Raman measurements, and Aria Khalili and Jae-Young Cho at NRC-QN for the AFM measurements. These data are not included in the manuscript. We thank Erika Janitz and Sigurd Fl{\aa}gan for useful discussions.
\end{acknowledgments}

\section*{Data Availability Statement}The data that support the findings of this study are available from the corresponding author upon reasonable request.

%\bibliography{annealingPaper}

\begin{thebibliography}{49}%
\makeatletter
\providecommand \@ifxundefined [1]{%
 \@ifx{#1\undefined}
}%
\providecommand \@ifnum [1]{%
 \ifnum #1\expandafter \@firstoftwo
 \else \expandafter \@secondoftwo
 \fi
}%
\providecommand \@ifx [1]{%
 \ifx #1\expandafter \@firstoftwo
 \else \expandafter \@secondoftwo
 \fi
}%
\providecommand \natexlab [1]{#1}%
\providecommand \enquote  [1]{``#1''}%
\providecommand \bibnamefont  [1]{#1}%
\providecommand \bibfnamefont [1]{#1}%
\providecommand \citenamefont [1]{#1}%
\providecommand \href@noop [0]{\@secondoftwo}%
\providecommand \href [0]{\begingroup \@sanitize@url \@href}%
\providecommand \@href[1]{\@@startlink{#1}\@@href}%
\providecommand \@@href[1]{\endgroup#1\@@endlink}%
\providecommand \@sanitize@url [0]{\catcode `\\12\catcode `\$12\catcode `\&12\catcode `\#12\catcode `\^12\catcode `\_12\catcode `\%12\relax}%
\providecommand \@@startlink[1]{}%
\providecommand \@@endlink[0]{}%
\providecommand \url  [0]{\begingroup\@sanitize@url \@url }%
\providecommand \@url [1]{\endgroup\@href {#1}{\urlprefix }}%
\providecommand \urlprefix  [0]{URL }%
\providecommand \Eprint [0]{\href }%
\providecommand \doibase [0]{https://doi.org/}%
\providecommand \selectlanguage [0]{\@gobble}%
\providecommand \bibinfo  [0]{\@secondoftwo}%
\providecommand \bibfield  [0]{\@secondoftwo}%
\providecommand \translation [1]{[#1]}%
\providecommand \BibitemOpen [0]{}%
\providecommand \bibitemStop [0]{}%
\providecommand \bibitemNoStop [0]{.\EOS\space}%
\providecommand \EOS [0]{\spacefactor3000\relax}%
\providecommand \BibitemShut  [1]{\csname bibitem#1\endcsname}%
\let\auto@bib@innerbib\@empty
%</preamble>
\bibitem [{\citenamefont {Dowling}\ and\ \citenamefont {Milburn}(2003)}]{ref:dowling2003qts}%
  \BibitemOpen
  \bibfield  {author} {\bibinfo {author} {\bibfnamefont {J.~P.}\ \bibnamefont {Dowling}}\ and\ \bibinfo {author} {\bibfnamefont {G.~J.}\ \bibnamefont {Milburn}},\ }\bibfield  {title} {\enquote {\bibinfo {title} {Quantum technology: The second quantum revolution},}\ }\href@noop {} {\bibfield  {journal} {\bibinfo  {journal} {Phil.~Trans.~R.~Soc.~A}\ }\textbf {\bibinfo {volume} {361}},\ \bibinfo {pages} {1655} (\bibinfo {year} {2003})}\BibitemShut {NoStop}%
\bibitem [{\citenamefont {Wang}\ \emph {et~al.}(2019)\citenamefont {Wang}, \citenamefont {Sciarrino}, \citenamefont {Laing},\ and\ \citenamefont {Thompson}}]{Wang-NaturePhys-2019-IntegratedQuantumTechnologies}%
  \BibitemOpen
  \bibfield  {author} {\bibinfo {author} {\bibfnamefont {J.}~\bibnamefont {Wang}}, \bibinfo {author} {\bibfnamefont {F.}~\bibnamefont {Sciarrino}}, \bibinfo {author} {\bibfnamefont {A.}~\bibnamefont {Laing}},\ and\ \bibinfo {author} {\bibfnamefont {M.~G.}\ \bibnamefont {Thompson}},\ }\bibfield  {title} {\enquote {\bibinfo {title} {{Integrated photonic quantum technologies}},}\ }\href {https://doi.org/10.1038/s41566-019-0532-1} {\bibfield  {journal} {\bibinfo  {journal} {Nature Photonics}\ ,\ \bibinfo {pages} {1--12}} (\bibinfo {year} {2019})}\BibitemShut {NoStop}%
\bibitem [{\citenamefont {Awschalom}\ \emph {et~al.}(2021)\citenamefont {Awschalom}, \citenamefont {Berggren}, \citenamefont {Bernien}, \citenamefont {Bhave}, \citenamefont {Carr}, \citenamefont {Davids}, \citenamefont {Economou}, \citenamefont {Englund}, \citenamefont {Faraon}, \citenamefont {Fejer} \emph {et~al.}}]{awschalom2021development}%
  \BibitemOpen
  \bibfield  {author} {\bibinfo {author} {\bibfnamefont {D.}~\bibnamefont {Awschalom}}, \bibinfo {author} {\bibfnamefont {K.~K.}\ \bibnamefont {Berggren}}, \bibinfo {author} {\bibfnamefont {H.}~\bibnamefont {Bernien}}, \bibinfo {author} {\bibfnamefont {S.}~\bibnamefont {Bhave}}, \bibinfo {author} {\bibfnamefont {L.~D.}\ \bibnamefont {Carr}}, \bibinfo {author} {\bibfnamefont {P.}~\bibnamefont {Davids}}, \bibinfo {author} {\bibfnamefont {S.~E.}\ \bibnamefont {Economou}}, \bibinfo {author} {\bibfnamefont {D.}~\bibnamefont {Englund}}, \bibinfo {author} {\bibfnamefont {A.}~\bibnamefont {Faraon}}, \bibinfo {author} {\bibfnamefont {M.}~\bibnamefont {Fejer}}, \emph {et~al.},\ }\bibfield  {title} {\enquote {\bibinfo {title} {Development of quantum interconnects (quics) for next-generation information technologies},}\ }\href@noop {} {\bibfield  {journal} {\bibinfo  {journal} {Prx Quantum}\ }\textbf {\bibinfo {volume} {2}},\ \bibinfo {pages} {017002} (\bibinfo {year} {2021})}\BibitemShut {NoStop}%
\bibitem [{\citenamefont {Monroe}\ \emph {et~al.}(2014)\citenamefont {Monroe}, \citenamefont {Raussendorf}, \citenamefont {Ruthven}, \citenamefont {Brown}, \citenamefont {Maunz}, \citenamefont {Duan},\ and\ \citenamefont {Kim}}]{Monroe-PRA-2014-ModularQuantumComputer}%
  \BibitemOpen
  \bibfield  {author} {\bibinfo {author} {\bibfnamefont {C.}~\bibnamefont {Monroe}}, \bibinfo {author} {\bibfnamefont {R.}~\bibnamefont {Raussendorf}}, \bibinfo {author} {\bibfnamefont {A.}~\bibnamefont {Ruthven}}, \bibinfo {author} {\bibfnamefont {K.~R.}\ \bibnamefont {Brown}}, \bibinfo {author} {\bibfnamefont {P.}~\bibnamefont {Maunz}}, \bibinfo {author} {\bibfnamefont {L.~M.}\ \bibnamefont {Duan}},\ and\ \bibinfo {author} {\bibfnamefont {J.}~\bibnamefont {Kim}},\ }\bibfield  {title} {\enquote {\bibinfo {title} {{Large-scale modular quantum-computer architecture with atomic memory and photonic interconnects}},}\ }\href {https://doi.org/10.1103/PhysRevA.89.022317} {\bibfield  {journal} {\bibinfo  {journal} {Physical Review A - Atomic, Molecular, and Optical Physics}\ }\textbf {\bibinfo {volume} {89}},\ \bibinfo {pages} {022317} (\bibinfo {year} {2014})},\ \Eprint {https://arxiv.org/abs/1208.0391} {arXiv:1208.0391} \BibitemShut {NoStop}%
\bibitem [{\citenamefont {Muralidharan}\ \emph {et~al.}(2016)\citenamefont {Muralidharan}, \citenamefont {Li}, \citenamefont {Kim}, \citenamefont {L{\"u}tkenhaus}, \citenamefont {Lukin},\ and\ \citenamefont {Jiang}}]{muralidharan2016optimal}%
  \BibitemOpen
  \bibfield  {author} {\bibinfo {author} {\bibfnamefont {S.}~\bibnamefont {Muralidharan}}, \bibinfo {author} {\bibfnamefont {L.}~\bibnamefont {Li}}, \bibinfo {author} {\bibfnamefont {J.}~\bibnamefont {Kim}}, \bibinfo {author} {\bibfnamefont {N.}~\bibnamefont {L{\"u}tkenhaus}}, \bibinfo {author} {\bibfnamefont {M.~D.}\ \bibnamefont {Lukin}},\ and\ \bibinfo {author} {\bibfnamefont {L.}~\bibnamefont {Jiang}},\ }\bibfield  {title} {\enquote {\bibinfo {title} {Optimal architectures for long distance quantum communication},}\ }\href@noop {} {\bibfield  {journal} {\bibinfo  {journal} {Scientific reports}\ }\textbf {\bibinfo {volume} {6}},\ \bibinfo {pages} {20463} (\bibinfo {year} {2016})}\BibitemShut {NoStop}%
\bibitem [{\citenamefont {K{\'{o}}m{\'{a}}r}\ \emph {et~al.}(2014)\citenamefont {K{\'{o}}m{\'{a}}r}, \citenamefont {Kessler}, \citenamefont {Bishof}, \citenamefont {Jiang}, \citenamefont {S{\o}rensen}, \citenamefont {Ye},\ and\ \citenamefont {Lukin}}]{Komar2014}%
  \BibitemOpen
  \bibfield  {author} {\bibinfo {author} {\bibfnamefont {P.}~\bibnamefont {K{\'{o}}m{\'{a}}r}}, \bibinfo {author} {\bibfnamefont {E.~M.}\ \bibnamefont {Kessler}}, \bibinfo {author} {\bibfnamefont {M.}~\bibnamefont {Bishof}}, \bibinfo {author} {\bibfnamefont {L.}~\bibnamefont {Jiang}}, \bibinfo {author} {\bibfnamefont {A.~S.}\ \bibnamefont {S{\o}rensen}}, \bibinfo {author} {\bibfnamefont {J.}~\bibnamefont {Ye}},\ and\ \bibinfo {author} {\bibfnamefont {M.~D.}\ \bibnamefont {Lukin}},\ }\bibfield  {title} {\enquote {\bibinfo {title} {{A quantum network of clocks}},}\ }\href {https://doi.org/10.1038/nphys3000} {\bibfield  {journal} {\bibinfo  {journal} {Nature Physics}\ }\textbf {\bibinfo {volume} {10}},\ \bibinfo {pages} {582--587} (\bibinfo {year} {2014})},\ \Eprint {https://arxiv.org/abs/1310.6045} {arXiv:1310.6045} \BibitemShut {NoStop}%
\bibitem [{\citenamefont {Kim}\ \emph {et~al.}(2023)\citenamefont {Kim}, \citenamefont {Choi}, \citenamefont {Trusheim}, \citenamefont {Wang},\ and\ \citenamefont {Englund}}]{kim2023nanophotonic}%
  \BibitemOpen
  \bibfield  {author} {\bibinfo {author} {\bibfnamefont {L.}~\bibnamefont {Kim}}, \bibinfo {author} {\bibfnamefont {H.}~\bibnamefont {Choi}}, \bibinfo {author} {\bibfnamefont {M.~E.}\ \bibnamefont {Trusheim}}, \bibinfo {author} {\bibfnamefont {H.}~\bibnamefont {Wang}},\ and\ \bibinfo {author} {\bibfnamefont {D.~R.}\ \bibnamefont {Englund}},\ }\bibfield  {title} {\enquote {\bibinfo {title} {Nanophotonic quantum sensing with engineered spin-optic coupling},}\ }\href@noop {} {\bibfield  {journal} {\bibinfo  {journal} {Nanophotonics}\ }\textbf {\bibinfo {volume} {12}},\ \bibinfo {pages} {441--449} (\bibinfo {year} {2023})}\BibitemShut {NoStop}%
\bibitem [{\citenamefont {Kimble}(2008)}]{Kimble2008}%
  \BibitemOpen
  \bibfield  {author} {\bibinfo {author} {\bibfnamefont {H.~J.}\ \bibnamefont {Kimble}},\ }\bibfield  {title} {\enquote {\bibinfo {title} {{The quantum internet}},}\ }\href {https://doi.org/10.1038/nature07127} {\bibfield  {journal} {\bibinfo  {journal} {Nature}\ }\textbf {\bibinfo {volume} {453}},\ \bibinfo {pages} {1023--1030} (\bibinfo {year} {2008})},\ \Eprint {https://arxiv.org/abs/0806.4195} {arXiv:0806.4195} \BibitemShut {NoStop}%
\bibitem [{\citenamefont {Janitz}, \citenamefont {Bhaskar},\ and\ \citenamefont {Childress}(2020)}]{Janitz2020}%
  \BibitemOpen
  \bibfield  {author} {\bibinfo {author} {\bibfnamefont {E.}~\bibnamefont {Janitz}}, \bibinfo {author} {\bibfnamefont {M.~K.}\ \bibnamefont {Bhaskar}},\ and\ \bibinfo {author} {\bibfnamefont {L.}~\bibnamefont {Childress}},\ }\bibfield  {title} {\enquote {\bibinfo {title} {Cavity quantum electrodynamics with color centers in diamond},}\ }\href {https://doi.org/10.1364/OPTICA.398628} {\bibfield  {journal} {\bibinfo  {journal} {Optica}\ }\textbf {\bibinfo {volume} {7}},\ \bibinfo {pages} {1232} (\bibinfo {year} {2020})}\BibitemShut {NoStop}%
\bibitem [{\citenamefont {Shandilya}\ \emph {et~al.}(2022)\citenamefont {Shandilya}, \citenamefont {Fl{\aa}gan}, \citenamefont {Carvalho}, \citenamefont {Zohari}, \citenamefont {Kavatamane}, \citenamefont {Losby},\ and\ \citenamefont {Barclay}}]{shandilya2022diamond}%
  \BibitemOpen
  \bibfield  {author} {\bibinfo {author} {\bibfnamefont {P.~K.}\ \bibnamefont {Shandilya}}, \bibinfo {author} {\bibfnamefont {S.}~\bibnamefont {Fl{\aa}gan}}, \bibinfo {author} {\bibfnamefont {N.~C.}\ \bibnamefont {Carvalho}}, \bibinfo {author} {\bibfnamefont {E.}~\bibnamefont {Zohari}}, \bibinfo {author} {\bibfnamefont {V.~K.}\ \bibnamefont {Kavatamane}}, \bibinfo {author} {\bibfnamefont {J.~E.}\ \bibnamefont {Losby}},\ and\ \bibinfo {author} {\bibfnamefont {P.~E.}\ \bibnamefont {Barclay}},\ }\bibfield  {title} {\enquote {\bibinfo {title} {Diamond integrated quantum nanophotonics: spins, photons and phonons},}\ }\href@noop {} {\bibfield  {journal} {\bibinfo  {journal} {Journal of Lightwave Technology}\ }\textbf {\bibinfo {volume} {40}},\ \bibinfo {pages} {7538--7571} (\bibinfo {year} {2022})}\BibitemShut {NoStop}%
\bibitem [{\citenamefont {Bar-Gill}\ \emph {et~al.}(2013)\citenamefont {Bar-Gill}, \citenamefont {Pham}, \citenamefont {Jarmola}, \citenamefont {Budker},\ and\ \citenamefont {Walsworth}}]{Bar-Gill2013}%
  \BibitemOpen
  \bibfield  {author} {\bibinfo {author} {\bibfnamefont {N.}~\bibnamefont {Bar-Gill}}, \bibinfo {author} {\bibfnamefont {L.}~\bibnamefont {Pham}}, \bibinfo {author} {\bibfnamefont {A.}~\bibnamefont {Jarmola}}, \bibinfo {author} {\bibfnamefont {D.}~\bibnamefont {Budker}},\ and\ \bibinfo {author} {\bibfnamefont {R.}~\bibnamefont {Walsworth}},\ }\bibfield  {title} {\enquote {\bibinfo {title} {{Solid-state electronic spin coherence time approaching one second}},}\ }\href {https://doi.org/10.1038/ncomms2771} {\bibfield  {journal} {\bibinfo  {journal} {Nature Communications}\ }\textbf {\bibinfo {volume} {4}},\ \bibinfo {pages} {1743} (\bibinfo {year} {2013})}\BibitemShut {NoStop}%
\bibitem [{\citenamefont {Bradley}\ \emph {et~al.}(2019)\citenamefont {Bradley}, \citenamefont {Randall}, \citenamefont {Abobeih}, \citenamefont {Berrevoets}, \citenamefont {Degen}, \citenamefont {Bakker}, \citenamefont {Markham}, \citenamefont {Twitchen},\ and\ \citenamefont {Taminiau}}]{Bradly-PRX-2019-NV10qubit1minuteCoherence}%
  \BibitemOpen
  \bibfield  {author} {\bibinfo {author} {\bibfnamefont {C.~E.}\ \bibnamefont {Bradley}}, \bibinfo {author} {\bibfnamefont {J.}~\bibnamefont {Randall}}, \bibinfo {author} {\bibfnamefont {M.~H.}\ \bibnamefont {Abobeih}}, \bibinfo {author} {\bibfnamefont {R.~C.}\ \bibnamefont {Berrevoets}}, \bibinfo {author} {\bibfnamefont {M.~J.}\ \bibnamefont {Degen}}, \bibinfo {author} {\bibfnamefont {M.~A.}\ \bibnamefont {Bakker}}, \bibinfo {author} {\bibfnamefont {M.}~\bibnamefont {Markham}}, \bibinfo {author} {\bibfnamefont {D.~J.}\ \bibnamefont {Twitchen}},\ and\ \bibinfo {author} {\bibfnamefont {T.~H.}\ \bibnamefont {Taminiau}},\ }\bibfield  {title} {\enquote {\bibinfo {title} {A ten-qubit solid-state spin register with quantum memory up to one minute},}\ }\href {https://doi.org/10.1103/PhysRevX.9.031045} {\bibfield  {journal} {\bibinfo  {journal} {Physical Review X}\ }\textbf {\bibinfo {volume} {9}},\ \bibinfo {pages} {031045} (\bibinfo {year} {2019})}\BibitemShut {NoStop}%
\bibitem [{\citenamefont {Myers}\ \emph {et~al.}(2014)\citenamefont {Myers}, \citenamefont {Das}, \citenamefont {Dartiailh}, \citenamefont {Ohno}, \citenamefont {Awschalom},\ and\ \citenamefont {Jayich}}]{myers2014probing}%
  \BibitemOpen
  \bibfield  {author} {\bibinfo {author} {\bibfnamefont {B.~A.}\ \bibnamefont {Myers}}, \bibinfo {author} {\bibfnamefont {A.}~\bibnamefont {Das}}, \bibinfo {author} {\bibfnamefont {M.}~\bibnamefont {Dartiailh}}, \bibinfo {author} {\bibfnamefont {K.}~\bibnamefont {Ohno}}, \bibinfo {author} {\bibfnamefont {D.~D.}\ \bibnamefont {Awschalom}},\ and\ \bibinfo {author} {\bibfnamefont {A.~B.}\ \bibnamefont {Jayich}},\ }\bibfield  {title} {\enquote {\bibinfo {title} {Probing surface noise with depth-calibrated spins in diamond},}\ }\href {https://doi.org/10.1103/PhysRevLett.113.027602} {\bibfield  {journal} {\bibinfo  {journal} {Physical Review Letters}\ }\textbf {\bibinfo {volume} {113}},\ \bibinfo {pages} {027602} (\bibinfo {year} {2014})}\BibitemShut {NoStop}%
\bibitem [{\citenamefont {Janitz}\ \emph {et~al.}(2022)\citenamefont {Janitz}, \citenamefont {Herb}, \citenamefont {V{\"{o}}lker}, \citenamefont {Huxter}, \citenamefont {Degen},\ and\ \citenamefont {Abendroth}}]{Janitz2022}%
  \BibitemOpen
  \bibfield  {author} {\bibinfo {author} {\bibfnamefont {E.}~\bibnamefont {Janitz}}, \bibinfo {author} {\bibfnamefont {K.}~\bibnamefont {Herb}}, \bibinfo {author} {\bibfnamefont {L.~A.}\ \bibnamefont {V{\"{o}}lker}}, \bibinfo {author} {\bibfnamefont {W.~S.}\ \bibnamefont {Huxter}}, \bibinfo {author} {\bibfnamefont {C.~L.}\ \bibnamefont {Degen}},\ and\ \bibinfo {author} {\bibfnamefont {J.~M.}\ \bibnamefont {Abendroth}},\ }\bibfield  {title} {\enquote {\bibinfo {title} {{Diamond surface engineering for molecular sensing with nitrogen—vacancy centers}},}\ }\href {https://doi.org/10.1039/D2TC01258H} {\bibfield  {journal} {\bibinfo  {journal} {Journal of Materials Chemistry C}\ } (\bibinfo {year} {2022}),\ 10.1039/D2TC01258H}\BibitemShut {NoStop}%
\bibitem [{\citenamefont {Hensen}\ \emph {et~al.}(2015)\citenamefont {Hensen}, \citenamefont {Bernien}, \citenamefont {Dr{\'{e}}au}, \citenamefont {Reiserer}, \citenamefont {Kalb}, \citenamefont {Blok}, \citenamefont {Ruitenberg}, \citenamefont {Vermeulen}, \citenamefont {Schouten}, \citenamefont {Abell{\'{a}}n}, \citenamefont {Amaya}, \citenamefont {Pruneri}, \citenamefont {Mitchell}, \citenamefont {Markham}, \citenamefont {Twitchen}, \citenamefont {Elkouss}, \citenamefont {Wehner}, \citenamefont {Taminiau},\ and\ \citenamefont {Hanson}}]{Hensen2015}%
  \BibitemOpen
  \bibfield  {author} {\bibinfo {author} {\bibfnamefont {B.}~\bibnamefont {Hensen}}, \bibinfo {author} {\bibfnamefont {H.}~\bibnamefont {Bernien}}, \bibinfo {author} {\bibfnamefont {A.~E.}\ \bibnamefont {Dr{\'{e}}au}}, \bibinfo {author} {\bibfnamefont {A.}~\bibnamefont {Reiserer}}, \bibinfo {author} {\bibfnamefont {N.}~\bibnamefont {Kalb}}, \bibinfo {author} {\bibfnamefont {M.~S.}\ \bibnamefont {Blok}}, \bibinfo {author} {\bibfnamefont {J.}~\bibnamefont {Ruitenberg}}, \bibinfo {author} {\bibfnamefont {R.~F.~L.}\ \bibnamefont {Vermeulen}}, \bibinfo {author} {\bibfnamefont {R.~N.}\ \bibnamefont {Schouten}}, \bibinfo {author} {\bibfnamefont {C.}~\bibnamefont {Abell{\'{a}}n}}, \bibinfo {author} {\bibfnamefont {W.}~\bibnamefont {Amaya}}, \bibinfo {author} {\bibfnamefont {V.}~\bibnamefont {Pruneri}}, \bibinfo {author} {\bibfnamefont {M.~W.}\ \bibnamefont {Mitchell}}, \bibinfo {author} {\bibfnamefont {M.}~\bibnamefont {Markham}}, \bibinfo {author} {\bibfnamefont {D.~J.}\ \bibnamefont {Twitchen}}, \bibinfo {author}
  {\bibfnamefont {D.}~\bibnamefont {Elkouss}}, \bibinfo {author} {\bibfnamefont {S.}~\bibnamefont {Wehner}}, \bibinfo {author} {\bibfnamefont {T.~H.}\ \bibnamefont {Taminiau}},\ and\ \bibinfo {author} {\bibfnamefont {R.}~\bibnamefont {Hanson}},\ }\bibfield  {title} {\enquote {\bibinfo {title} {{Experimental loophole-free violation of a Bell inequality using entangled electron spins separated by 1.3 km}},}\ }\href {http://arxiv.org/abs/1508.05949} {\  (\bibinfo {year} {2015})},\ \Eprint {https://arxiv.org/abs/1508.05949} {arXiv:1508.05949} \BibitemShut {NoStop}%
\bibitem [{\citenamefont {Bradac}\ \emph {et~al.}(2019)\citenamefont {Bradac}, \citenamefont {Gao}, \citenamefont {Forneris}, \citenamefont {Trusheim},\ and\ \citenamefont {Aharonovich}}]{Bradac2019}%
  \BibitemOpen
  \bibfield  {author} {\bibinfo {author} {\bibfnamefont {C.}~\bibnamefont {Bradac}}, \bibinfo {author} {\bibfnamefont {W.}~\bibnamefont {Gao}}, \bibinfo {author} {\bibfnamefont {J.}~\bibnamefont {Forneris}}, \bibinfo {author} {\bibfnamefont {M.~E.}\ \bibnamefont {Trusheim}},\ and\ \bibinfo {author} {\bibfnamefont {I.}~\bibnamefont {Aharonovich}},\ }\bibfield  {title} {\enquote {\bibinfo {title} {Quantum nanophotonics with group iv defects in diamond},}\ }\href {https://doi.org/10.1038/s41467-019-13332-w} {\bibfield  {journal} {\bibinfo  {journal} {Nature Communications}\ }\textbf {\bibinfo {volume} {10}},\ \bibinfo {pages} {5625} (\bibinfo {year} {2019})}\BibitemShut {NoStop}%
\bibitem [{\citenamefont {Nguyen}\ \emph {et~al.}(2019)\citenamefont {Nguyen}, \citenamefont {Sukachev}, \citenamefont {Bhaskar}, \citenamefont {Machielse}, \citenamefont {Levonian}, \citenamefont {Knall}, \citenamefont {Stroganov}, \citenamefont {Chia}, \citenamefont {Burek}, \citenamefont {Riedinger}, \citenamefont {Park}, \citenamefont {Lon\v{c}ar},\ and\ \citenamefont {Lukin}}]{Nguyen2019}%
  \BibitemOpen
  \bibfield  {author} {\bibinfo {author} {\bibfnamefont {C.~T.}\ \bibnamefont {Nguyen}}, \bibinfo {author} {\bibfnamefont {D.~D.}\ \bibnamefont {Sukachev}}, \bibinfo {author} {\bibfnamefont {M.~K.}\ \bibnamefont {Bhaskar}}, \bibinfo {author} {\bibfnamefont {B.}~\bibnamefont {Machielse}}, \bibinfo {author} {\bibfnamefont {D.~S.}\ \bibnamefont {Levonian}}, \bibinfo {author} {\bibfnamefont {E.~N.}\ \bibnamefont {Knall}}, \bibinfo {author} {\bibfnamefont {P.}~\bibnamefont {Stroganov}}, \bibinfo {author} {\bibfnamefont {C.}~\bibnamefont {Chia}}, \bibinfo {author} {\bibfnamefont {M.~J.}\ \bibnamefont {Burek}}, \bibinfo {author} {\bibfnamefont {R.}~\bibnamefont {Riedinger}}, \bibinfo {author} {\bibfnamefont {H.}~\bibnamefont {Park}}, \bibinfo {author} {\bibfnamefont {M.}~\bibnamefont {Lon\v{c}ar}},\ and\ \bibinfo {author} {\bibfnamefont {M.~D.}\ \bibnamefont {Lukin}},\ }\bibfield  {title} {\enquote {\bibinfo {title} {An integrated nanophotonic quantum register based on silicon-vacancy spins in diamond},}\ }\href
  {https://doi.org/10.1103/PhysRevB.100.165428} {\bibfield  {journal} {\bibinfo  {journal} {Physical Review B}\ }\textbf {\bibinfo {volume} {100}},\ \bibinfo {pages} {165428} (\bibinfo {year} {2019})}\BibitemShut {NoStop}%
\bibitem [{\citenamefont {Pompili}\ \emph {et~al.}(2021)\citenamefont {Pompili}, \citenamefont {Hermans}, \citenamefont {Baier}, \citenamefont {Beukers}, \citenamefont {Humphreys}, \citenamefont {Schouten}, \citenamefont {Vermeulen}, \citenamefont {Tiggelman}, \citenamefont {dos Santos~Martins}, \citenamefont {Dirkse} \emph {et~al.}}]{pompili2021realization}%
  \BibitemOpen
  \bibfield  {author} {\bibinfo {author} {\bibfnamefont {M.}~\bibnamefont {Pompili}}, \bibinfo {author} {\bibfnamefont {S.~L.}\ \bibnamefont {Hermans}}, \bibinfo {author} {\bibfnamefont {S.}~\bibnamefont {Baier}}, \bibinfo {author} {\bibfnamefont {H.~K.}\ \bibnamefont {Beukers}}, \bibinfo {author} {\bibfnamefont {P.~C.}\ \bibnamefont {Humphreys}}, \bibinfo {author} {\bibfnamefont {R.~N.}\ \bibnamefont {Schouten}}, \bibinfo {author} {\bibfnamefont {R.~F.}\ \bibnamefont {Vermeulen}}, \bibinfo {author} {\bibfnamefont {M.~J.}\ \bibnamefont {Tiggelman}}, \bibinfo {author} {\bibfnamefont {L.}~\bibnamefont {dos Santos~Martins}}, \bibinfo {author} {\bibfnamefont {B.}~\bibnamefont {Dirkse}}, \emph {et~al.},\ }\bibfield  {title} {\enquote {\bibinfo {title} {Realization of a multinode quantum network of remote solid-state qubits},}\ }\href {http://doi.org/10.1126/science.abg1919} {\bibfield  {journal} {\bibinfo  {journal} {Science}\ }\textbf {\bibinfo {volume} {372}},\ \bibinfo {pages} {259--264} (\bibinfo {year}
  {2021})}\BibitemShut {NoStop}%
\bibitem [{\citenamefont {Hermans}\ \emph {et~al.}(2022)\citenamefont {Hermans}, \citenamefont {Pompili}, \citenamefont {Beukers}, \citenamefont {Baier}, \citenamefont {Borregaard},\ and\ \citenamefont {Hanson}}]{Hermans2022}%
  \BibitemOpen
  \bibfield  {author} {\bibinfo {author} {\bibfnamefont {S.~L.~N.}\ \bibnamefont {Hermans}}, \bibinfo {author} {\bibfnamefont {M.}~\bibnamefont {Pompili}}, \bibinfo {author} {\bibfnamefont {H.~K.~C.}\ \bibnamefont {Beukers}}, \bibinfo {author} {\bibfnamefont {S.}~\bibnamefont {Baier}}, \bibinfo {author} {\bibfnamefont {J.}~\bibnamefont {Borregaard}},\ and\ \bibinfo {author} {\bibfnamefont {R.}~\bibnamefont {Hanson}},\ }\bibfield  {title} {\enquote {\bibinfo {title} {Qubit teleportation between non-neighbouring nodes in a quantum network},}\ }\href {https://doi.org/10.1038/s41586-022-04697-y} {\bibfield  {journal} {\bibinfo  {journal} {Nature}\ }\textbf {\bibinfo {volume} {605}},\ \bibinfo {pages} {663--668} (\bibinfo {year} {2022})}\BibitemShut {NoStop}%
\bibitem [{\citenamefont {Knaut}\ \emph {et~al.}(2024)\citenamefont {Knaut}, \citenamefont {Suleymanzade}, \citenamefont {Wei}, \citenamefont {Assumpcao}, \citenamefont {Stas}, \citenamefont {Huan}, \citenamefont {Machielse}, \citenamefont {Knall}, \citenamefont {Sutula}, \citenamefont {Baranes} \emph {et~al.}}]{knaut2024entanglement}%
  \BibitemOpen
  \bibfield  {author} {\bibinfo {author} {\bibfnamefont {C.}~\bibnamefont {Knaut}}, \bibinfo {author} {\bibfnamefont {A.}~\bibnamefont {Suleymanzade}}, \bibinfo {author} {\bibfnamefont {Y.-C.}\ \bibnamefont {Wei}}, \bibinfo {author} {\bibfnamefont {D.}~\bibnamefont {Assumpcao}}, \bibinfo {author} {\bibfnamefont {P.-J.}\ \bibnamefont {Stas}}, \bibinfo {author} {\bibfnamefont {Y.}~\bibnamefont {Huan}}, \bibinfo {author} {\bibfnamefont {B.}~\bibnamefont {Machielse}}, \bibinfo {author} {\bibfnamefont {E.}~\bibnamefont {Knall}}, \bibinfo {author} {\bibfnamefont {M.}~\bibnamefont {Sutula}}, \bibinfo {author} {\bibfnamefont {G.}~\bibnamefont {Baranes}}, \emph {et~al.},\ }\bibfield  {title} {\enquote {\bibinfo {title} {Entanglement of nanophotonic quantum memory nodes in a telecom network},}\ }\href@noop {} {\bibfield  {journal} {\bibinfo  {journal} {Nature}\ }\textbf {\bibinfo {volume} {629}},\ \bibinfo {pages} {573--578} (\bibinfo {year} {2024})}\BibitemShut {NoStop}%
\bibitem [{\citenamefont {Faraon}\ \emph {et~al.}(2013)\citenamefont {Faraon}, \citenamefont {Santori}, \citenamefont {Huang}, \citenamefont {Fu}, \citenamefont {Acosta}, \citenamefont {Fattal},\ and\ \citenamefont {Beausoleil}}]{Faraon2013}%
  \BibitemOpen
  \bibfield  {author} {\bibinfo {author} {\bibfnamefont {A.}~\bibnamefont {Faraon}}, \bibinfo {author} {\bibfnamefont {C.}~\bibnamefont {Santori}}, \bibinfo {author} {\bibfnamefont {Z.}~\bibnamefont {Huang}}, \bibinfo {author} {\bibfnamefont {K.~M.~C.}\ \bibnamefont {Fu}}, \bibinfo {author} {\bibfnamefont {V.~M.}\ \bibnamefont {Acosta}}, \bibinfo {author} {\bibfnamefont {D.}~\bibnamefont {Fattal}},\ and\ \bibinfo {author} {\bibfnamefont {R.~G.}\ \bibnamefont {Beausoleil}},\ }\bibfield  {title} {\enquote {\bibinfo {title} {{Quantum photonic devices in single-crystal diamond}},}\ }\href {https://doi.org/10.1088/1367-2630/15/2/025010} {\bibfield  {journal} {\bibinfo  {journal} {New Journal of Physics}\ }\textbf {\bibinfo {volume} {15}},\ \bibinfo {pages} {025010} (\bibinfo {year} {2013})}\BibitemShut {NoStop}%
\bibitem [{\citenamefont {Schr{\"{o}}der}\ \emph {et~al.}(2017)\citenamefont {Schr{\"{o}}der}, \citenamefont {Trusheim}, \citenamefont {Walsh}, \citenamefont {Li}, \citenamefont {Zheng}, \citenamefont {Schukraft}, \citenamefont {Sipahigil}, \citenamefont {Evans}, \citenamefont {Sukachev}, \citenamefont {Nguyen}, \citenamefont {Pacheco}, \citenamefont {Camacho}, \citenamefont {Bielejec}, \citenamefont {Lukin},\ and\ \citenamefont {Englund}}]{Schroder2017}%
  \BibitemOpen
  \bibfield  {author} {\bibinfo {author} {\bibfnamefont {T.}~\bibnamefont {Schr{\"{o}}der}}, \bibinfo {author} {\bibfnamefont {M.~E.}\ \bibnamefont {Trusheim}}, \bibinfo {author} {\bibfnamefont {M.}~\bibnamefont {Walsh}}, \bibinfo {author} {\bibfnamefont {L.}~\bibnamefont {Li}}, \bibinfo {author} {\bibfnamefont {J.}~\bibnamefont {Zheng}}, \bibinfo {author} {\bibfnamefont {M.}~\bibnamefont {Schukraft}}, \bibinfo {author} {\bibfnamefont {A.}~\bibnamefont {Sipahigil}}, \bibinfo {author} {\bibfnamefont {R.~E.}\ \bibnamefont {Evans}}, \bibinfo {author} {\bibfnamefont {D.~D.}\ \bibnamefont {Sukachev}}, \bibinfo {author} {\bibfnamefont {C.~T.}\ \bibnamefont {Nguyen}}, \bibinfo {author} {\bibfnamefont {J.~L.}\ \bibnamefont {Pacheco}}, \bibinfo {author} {\bibfnamefont {R.~M.}\ \bibnamefont {Camacho}}, \bibinfo {author} {\bibfnamefont {E.~S.}\ \bibnamefont {Bielejec}}, \bibinfo {author} {\bibfnamefont {M.~D.}\ \bibnamefont {Lukin}},\ and\ \bibinfo {author} {\bibfnamefont {D.}~\bibnamefont {Englund}},\ }\bibfield
  {title} {\enquote {\bibinfo {title} {Scalable focused ion beam creation of nearly lifetime-limited single quantum emitters in diamond nanostructures},}\ }\href {https://doi.org/10.1038/ncomms15376} {\bibfield  {journal} {\bibinfo  {journal} {Nature Communications}\ }\textbf {\bibinfo {volume} {8}},\ \bibinfo {pages} {15376} (\bibinfo {year} {2017})}\BibitemShut {NoStop}%
\bibitem [{\citenamefont {Shandilya}\ \emph {et~al.}(2021)\citenamefont {Shandilya}, \citenamefont {Lake}, \citenamefont {Mitchell}, \citenamefont {Sukachev},\ and\ \citenamefont {Barclay}}]{shandilya2021optomechanical}%
  \BibitemOpen
  \bibfield  {author} {\bibinfo {author} {\bibfnamefont {P.~K.}\ \bibnamefont {Shandilya}}, \bibinfo {author} {\bibfnamefont {D.~P.}\ \bibnamefont {Lake}}, \bibinfo {author} {\bibfnamefont {M.~J.}\ \bibnamefont {Mitchell}}, \bibinfo {author} {\bibfnamefont {D.~D.}\ \bibnamefont {Sukachev}},\ and\ \bibinfo {author} {\bibfnamefont {P.~E.}\ \bibnamefont {Barclay}},\ }\bibfield  {title} {\enquote {\bibinfo {title} {Optomechanical interface between telecom photons and spin quantum memory},}\ }\href {https://doi.org/10.1038/s41567-021-01364-3} {\bibfield  {journal} {\bibinfo  {journal} {Nature Physics}\ }\textbf {\bibinfo {volume} {17}},\ \bibinfo {pages} {1420--1425} (\bibinfo {year} {2021})}\BibitemShut {NoStop}%
\bibitem [{\citenamefont {Li}\ \emph {et~al.}(2015)\citenamefont {Li}, \citenamefont {Schr\"{o}der}, \citenamefont {Chen}, \citenamefont {Walsh}, \citenamefont {Bayn}, \citenamefont {Goldstein}, \citenamefont {Gaathon}, \citenamefont {Trusheim}, \citenamefont {Lu}, \citenamefont {Mower}, \citenamefont {Cotlet}, \citenamefont {Markham}, \citenamefont {Twitchen},\ and\ \citenamefont {Englund}}]{Li2014}%
  \BibitemOpen
  \bibfield  {author} {\bibinfo {author} {\bibfnamefont {L.}~\bibnamefont {Li}}, \bibinfo {author} {\bibfnamefont {T.}~\bibnamefont {Schr\"{o}der}}, \bibinfo {author} {\bibfnamefont {E.~H.}\ \bibnamefont {Chen}}, \bibinfo {author} {\bibfnamefont {M.}~\bibnamefont {Walsh}}, \bibinfo {author} {\bibfnamefont {I.}~\bibnamefont {Bayn}}, \bibinfo {author} {\bibfnamefont {J.}~\bibnamefont {Goldstein}}, \bibinfo {author} {\bibfnamefont {O.}~\bibnamefont {Gaathon}}, \bibinfo {author} {\bibfnamefont {M.~E.}\ \bibnamefont {Trusheim}}, \bibinfo {author} {\bibfnamefont {M.}~\bibnamefont {Lu}}, \bibinfo {author} {\bibfnamefont {J.}~\bibnamefont {Mower}}, \bibinfo {author} {\bibfnamefont {M.}~\bibnamefont {Cotlet}}, \bibinfo {author} {\bibfnamefont {M.~L.}\ \bibnamefont {Markham}}, \bibinfo {author} {\bibfnamefont {D.~J.}\ \bibnamefont {Twitchen}},\ and\ \bibinfo {author} {\bibfnamefont {D.}~\bibnamefont {Englund}},\ }\bibfield  {title} {\enquote {\bibinfo {title} {Coherent spin control of a nanocavity-enhanced qubit in
  diamond},}\ }\href {https://doi.org/10.1038/ncomms7173} {\bibfield  {journal} {\bibinfo  {journal} {Nature Communications}\ }\textbf {\bibinfo {volume} {6}},\ \bibinfo {pages} {6173} (\bibinfo {year} {2015})}\BibitemShut {NoStop}%
\bibitem [{\citenamefont {Pezzagna}\ \emph {et~al.}(2011)\citenamefont {Pezzagna}, \citenamefont {Rogalla}, \citenamefont {Wildanger}, \citenamefont {Meijer},\ and\ \citenamefont {Zaitsev}}]{Pezzagna2011}%
  \BibitemOpen
  \bibfield  {author} {\bibinfo {author} {\bibfnamefont {S.}~\bibnamefont {Pezzagna}}, \bibinfo {author} {\bibfnamefont {D.}~\bibnamefont {Rogalla}}, \bibinfo {author} {\bibfnamefont {D.}~\bibnamefont {Wildanger}}, \bibinfo {author} {\bibfnamefont {J.}~\bibnamefont {Meijer}},\ and\ \bibinfo {author} {\bibfnamefont {A.}~\bibnamefont {Zaitsev}},\ }\bibfield  {title} {\enquote {\bibinfo {title} {Creation and nature of optical centres in diamond for single-photon emission—overview and critical remarks},}\ }\href {https://doi.org/10.1088/1367-2630/13/3/035024} {\bibfield  {journal} {\bibinfo  {journal} {New Journal of Physics}\ }\textbf {\bibinfo {volume} {13}},\ \bibinfo {pages} {035024} (\bibinfo {year} {2011})}\BibitemShut {NoStop}%
\bibitem [{\citenamefont {Chu}\ \emph {et~al.}(2014)\citenamefont {Chu}, \citenamefont {de~Leon}, \citenamefont {Shields}, \citenamefont {Hausmann}, \citenamefont {Evans}, \citenamefont {Togan}, \citenamefont {Burek}, \citenamefont {Markham}, \citenamefont {Stacey}, \citenamefont {Zibrov}, \citenamefont {Yacoby}, \citenamefont {Twitchen}, \citenamefont {Loncar}, \citenamefont {Park}, \citenamefont {Maletinsky},\ and\ \citenamefont {Lukin}}]{Chu2014NanoLett}%
  \BibitemOpen
  \bibfield  {author} {\bibinfo {author} {\bibfnamefont {Y.}~\bibnamefont {Chu}}, \bibinfo {author} {\bibfnamefont {N.}~\bibnamefont {de~Leon}}, \bibinfo {author} {\bibfnamefont {B.}~\bibnamefont {Shields}}, \bibinfo {author} {\bibfnamefont {B.}~\bibnamefont {Hausmann}}, \bibinfo {author} {\bibfnamefont {R.}~\bibnamefont {Evans}}, \bibinfo {author} {\bibfnamefont {E.}~\bibnamefont {Togan}}, \bibinfo {author} {\bibfnamefont {M.~J.}\ \bibnamefont {Burek}}, \bibinfo {author} {\bibfnamefont {M.}~\bibnamefont {Markham}}, \bibinfo {author} {\bibfnamefont {A.}~\bibnamefont {Stacey}}, \bibinfo {author} {\bibfnamefont {A.}~\bibnamefont {Zibrov}}, \bibinfo {author} {\bibfnamefont {A.}~\bibnamefont {Yacoby}}, \bibinfo {author} {\bibfnamefont {D.}~\bibnamefont {Twitchen}}, \bibinfo {author} {\bibfnamefont {M.}~\bibnamefont {Loncar}}, \bibinfo {author} {\bibfnamefont {H.}~\bibnamefont {Park}}, \bibinfo {author} {\bibfnamefont {P.}~\bibnamefont {Maletinsky}},\ and\ \bibinfo {author} {\bibfnamefont {M.}~\bibnamefont
  {Lukin}},\ }\bibfield  {title} {\enquote {\bibinfo {title} {Coherent optical transitions in implanted nitrogen vacancy centers},}\ }\href {https://doi.org/10.1021/nl404836p} {\bibfield  {journal} {\bibinfo  {journal} {Nano Letters}\ }\textbf {\bibinfo {volume} {14}},\ \bibinfo {pages} {1982--1986} (\bibinfo {year} {2014})}\BibitemShut {NoStop}%
\bibitem [{\citenamefont {Sipahigil}\ \emph {et~al.}(2016)\citenamefont {Sipahigil}, \citenamefont {Evans}, \citenamefont {Sukachev}, \citenamefont {Burek}, \citenamefont {Borregaard}, \citenamefont {Bhaskar}, \citenamefont {Nguyen}, \citenamefont {Pacheco}, \citenamefont {Atikian}, \citenamefont {Meuwly}, \citenamefont {Camacho}, \citenamefont {Jelezko}, \citenamefont {Bielejec}, \citenamefont {Park}, \citenamefont {Lončar},\ and\ \citenamefont {Lukin}}]{Sipahigil-Science-2016-SiVplatform}%
  \BibitemOpen
  \bibfield  {author} {\bibinfo {author} {\bibfnamefont {A.}~\bibnamefont {Sipahigil}}, \bibinfo {author} {\bibfnamefont {R.~E.}\ \bibnamefont {Evans}}, \bibinfo {author} {\bibfnamefont {D.~D.}\ \bibnamefont {Sukachev}}, \bibinfo {author} {\bibfnamefont {M.~J.}\ \bibnamefont {Burek}}, \bibinfo {author} {\bibfnamefont {J.}~\bibnamefont {Borregaard}}, \bibinfo {author} {\bibfnamefont {M.~K.}\ \bibnamefont {Bhaskar}}, \bibinfo {author} {\bibfnamefont {C.~T.}\ \bibnamefont {Nguyen}}, \bibinfo {author} {\bibfnamefont {J.~L.}\ \bibnamefont {Pacheco}}, \bibinfo {author} {\bibfnamefont {H.~A.}\ \bibnamefont {Atikian}}, \bibinfo {author} {\bibfnamefont {C.}~\bibnamefont {Meuwly}}, \bibinfo {author} {\bibfnamefont {R.~M.}\ \bibnamefont {Camacho}}, \bibinfo {author} {\bibfnamefont {F.}~\bibnamefont {Jelezko}}, \bibinfo {author} {\bibfnamefont {E.}~\bibnamefont {Bielejec}}, \bibinfo {author} {\bibfnamefont {H.}~\bibnamefont {Park}}, \bibinfo {author} {\bibfnamefont {M.}~\bibnamefont {Lončar}},\ and\ \bibinfo {author}
  {\bibfnamefont {M.~D.}\ \bibnamefont {Lukin}},\ }\bibfield  {title} {\enquote {\bibinfo {title} {An integrated diamond nanophotonics platform for quantum-optical networks},}\ }\href {https://doi.org/10.1126/science.aah6875} {\bibfield  {journal} {\bibinfo  {journal} {Science}\ }\textbf {\bibinfo {volume} {354}},\ \bibinfo {pages} {847--850} (\bibinfo {year} {2016})}\BibitemShut {NoStop}%
\bibitem [{\citenamefont {Wang}\ and\ \citenamefont {Lekavicius}(2020)}]{Wang2020APL}%
  \BibitemOpen
  \bibfield  {author} {\bibinfo {author} {\bibfnamefont {H.}~\bibnamefont {Wang}}\ and\ \bibinfo {author} {\bibfnamefont {I.}~\bibnamefont {Lekavicius}},\ }\bibfield  {title} {\enquote {\bibinfo {title} {Coupling spins to nanomechanical resonators: Toward quantum spin-mechanics},}\ }\href {https://doi.org/10.1063/5.0024001} {\bibfield  {journal} {\bibinfo  {journal} {Applied Physics Letters}\ }\textbf {\bibinfo {volume} {117}},\ \bibinfo {pages} {230501} (\bibinfo {year} {2020})}\BibitemShut {NoStop}%
\bibitem [{\citenamefont {Yamamoto}\ \emph {et~al.}(2013)\citenamefont {Yamamoto}, \citenamefont {Umeda}, \citenamefont {Watanabe}, \citenamefont {Onoda}, \citenamefont {Markham}, \citenamefont {Twitchen}, \citenamefont {Naydenov}, \citenamefont {McGuinness}, \citenamefont {Teraji}, \citenamefont {Koizumi}, \citenamefont {Dolde}, \citenamefont {Fedder}, \citenamefont {Honert}, \citenamefont {Wrachtrup}, \citenamefont {Ohshima}, \citenamefont {Jelezko},\ and\ \citenamefont {Isoya}}]{Yamamoto2013}%
  \BibitemOpen
  \bibfield  {author} {\bibinfo {author} {\bibfnamefont {T.}~\bibnamefont {Yamamoto}}, \bibinfo {author} {\bibfnamefont {T.}~\bibnamefont {Umeda}}, \bibinfo {author} {\bibfnamefont {K.}~\bibnamefont {Watanabe}}, \bibinfo {author} {\bibfnamefont {S.}~\bibnamefont {Onoda}}, \bibinfo {author} {\bibfnamefont {M.~L.}\ \bibnamefont {Markham}}, \bibinfo {author} {\bibfnamefont {D.~J.}\ \bibnamefont {Twitchen}}, \bibinfo {author} {\bibfnamefont {B.}~\bibnamefont {Naydenov}}, \bibinfo {author} {\bibfnamefont {L.~P.}\ \bibnamefont {McGuinness}}, \bibinfo {author} {\bibfnamefont {T.}~\bibnamefont {Teraji}}, \bibinfo {author} {\bibfnamefont {S.}~\bibnamefont {Koizumi}}, \bibinfo {author} {\bibfnamefont {F.}~\bibnamefont {Dolde}}, \bibinfo {author} {\bibfnamefont {H.}~\bibnamefont {Fedder}}, \bibinfo {author} {\bibfnamefont {J.}~\bibnamefont {Honert}}, \bibinfo {author} {\bibfnamefont {J.}~\bibnamefont {Wrachtrup}}, \bibinfo {author} {\bibfnamefont {T.}~\bibnamefont {Ohshima}}, \bibinfo {author} {\bibfnamefont
  {F.}~\bibnamefont {Jelezko}},\ and\ \bibinfo {author} {\bibfnamefont {J.}~\bibnamefont {Isoya}},\ }\bibfield  {title} {\enquote {\bibinfo {title} {{Extending spin coherence times of diamond qubits by high-temperature annealing}},}\ }\href {https://doi.org/10.1103/PhysRevB.88.075206} {\bibfield  {journal} {\bibinfo  {journal} {Physical Review B}\ }\textbf {\bibinfo {volume} {88}},\ \bibinfo {pages} {075206} (\bibinfo {year} {2013})}\BibitemShut {NoStop}%
\bibitem [{\citenamefont {Sangtawesin}\ \emph {et~al.}(2019)\citenamefont {Sangtawesin}, \citenamefont {Dwyer}, \citenamefont {Srinivasan}, \citenamefont {Allred}, \citenamefont {Rodgers}, \citenamefont {Greve}, \citenamefont {Stacey}, \citenamefont {Dontschuk}, \citenamefont {O’Donnell}, \citenamefont {Hu}, \citenamefont {Evans}, \citenamefont {Jaye}, \citenamefont {Fischer}, \citenamefont {Markham}, \citenamefont {Twitchen}, \citenamefont {Park}, \citenamefont {Lukin},\ and\ \citenamefont {de~Leon}}]{Sangtawesin2019}%
  \BibitemOpen
  \bibfield  {author} {\bibinfo {author} {\bibfnamefont {S.}~\bibnamefont {Sangtawesin}}, \bibinfo {author} {\bibfnamefont {B.~L.}\ \bibnamefont {Dwyer}}, \bibinfo {author} {\bibfnamefont {S.}~\bibnamefont {Srinivasan}}, \bibinfo {author} {\bibfnamefont {J.~J.}\ \bibnamefont {Allred}}, \bibinfo {author} {\bibfnamefont {L.~V.}\ \bibnamefont {Rodgers}}, \bibinfo {author} {\bibfnamefont {K.~D.}\ \bibnamefont {Greve}}, \bibinfo {author} {\bibfnamefont {A.}~\bibnamefont {Stacey}}, \bibinfo {author} {\bibfnamefont {N.}~\bibnamefont {Dontschuk}}, \bibinfo {author} {\bibfnamefont {K.~M.}\ \bibnamefont {O’Donnell}}, \bibinfo {author} {\bibfnamefont {D.}~\bibnamefont {Hu}}, \bibinfo {author} {\bibfnamefont {D.~A.}\ \bibnamefont {Evans}}, \bibinfo {author} {\bibfnamefont {C.}~\bibnamefont {Jaye}}, \bibinfo {author} {\bibfnamefont {D.~A.}\ \bibnamefont {Fischer}}, \bibinfo {author} {\bibfnamefont {M.~L.}\ \bibnamefont {Markham}}, \bibinfo {author} {\bibfnamefont {D.~J.}\ \bibnamefont {Twitchen}}, \bibinfo {author}
  {\bibfnamefont {H.}~\bibnamefont {Park}}, \bibinfo {author} {\bibfnamefont {M.~D.}\ \bibnamefont {Lukin}},\ and\ \bibinfo {author} {\bibfnamefont {N.~P.}\ \bibnamefont {de~Leon}},\ }\bibfield  {title} {\enquote {\bibinfo {title} {Origins of diamond surface noise probed by correlating single-spin measurements with surface spectroscopy},}\ }\href {https://doi.org/10.1103/PhysRevX.9.031052} {\bibfield  {journal} {\bibinfo  {journal} {Physical Review X}\ }\textbf {\bibinfo {volume} {9}},\ \bibinfo {pages} {031052} (\bibinfo {year} {2019})}\BibitemShut {NoStop}%
\bibitem [{\citenamefont {Jantzen}\ \emph {et~al.}(2016)\citenamefont {Jantzen}, \citenamefont {Kurz}, \citenamefont {Rudnicki}, \citenamefont {Sch{\"a}fermeier}, \citenamefont {Jahnke}, \citenamefont {Andersen}, \citenamefont {Davydov}, \citenamefont {Agafonov}, \citenamefont {Kubanek}, \citenamefont {Rogers} \emph {et~al.}}]{jantzen2016nanodiamonds}%
  \BibitemOpen
  \bibfield  {author} {\bibinfo {author} {\bibfnamefont {U.}~\bibnamefont {Jantzen}}, \bibinfo {author} {\bibfnamefont {A.~B.}\ \bibnamefont {Kurz}}, \bibinfo {author} {\bibfnamefont {D.~S.}\ \bibnamefont {Rudnicki}}, \bibinfo {author} {\bibfnamefont {C.}~\bibnamefont {Sch{\"a}fermeier}}, \bibinfo {author} {\bibfnamefont {K.~D.}\ \bibnamefont {Jahnke}}, \bibinfo {author} {\bibfnamefont {U.~L.}\ \bibnamefont {Andersen}}, \bibinfo {author} {\bibfnamefont {V.~A.}\ \bibnamefont {Davydov}}, \bibinfo {author} {\bibfnamefont {V.~N.}\ \bibnamefont {Agafonov}}, \bibinfo {author} {\bibfnamefont {A.}~\bibnamefont {Kubanek}}, \bibinfo {author} {\bibfnamefont {L.~J.}\ \bibnamefont {Rogers}}, \emph {et~al.},\ }\bibfield  {title} {\enquote {\bibinfo {title} {Nanodiamonds carrying silicon-vacancy quantum emitters with almost lifetime-limited linewidths},}\ }\href@noop {} {\bibfield  {journal} {\bibinfo  {journal} {New Journal of Physics}\ }\textbf {\bibinfo {volume} {18}},\ \bibinfo {pages} {073036} (\bibinfo {year}
  {2016})}\BibitemShut {NoStop}%
\bibitem [{\citenamefont {Faraon}\ \emph {et~al.}(2012)\citenamefont {Faraon}, \citenamefont {Santori}, \citenamefont {Huang}, \citenamefont {Acosta},\ and\ \citenamefont {Beausoleil}}]{Faraon2012}%
  \BibitemOpen
  \bibfield  {author} {\bibinfo {author} {\bibfnamefont {A.}~\bibnamefont {Faraon}}, \bibinfo {author} {\bibfnamefont {C.}~\bibnamefont {Santori}}, \bibinfo {author} {\bibfnamefont {Z.}~\bibnamefont {Huang}}, \bibinfo {author} {\bibfnamefont {V.~M.}\ \bibnamefont {Acosta}},\ and\ \bibinfo {author} {\bibfnamefont {R.~G.}\ \bibnamefont {Beausoleil}},\ }\bibfield  {title} {\enquote {\bibinfo {title} {{Coupling of Nitrogen-Vacancy Centers to Photonic Crystal Cavities in Monocrystalline Diamond}},}\ }\href {https://doi.org/10.1103/PhysRevLett.109.033604} {\bibfield  {journal} {\bibinfo  {journal} {Physical Review Letters}\ }\textbf {\bibinfo {volume} {109}},\ \bibinfo {pages} {033604} (\bibinfo {year} {2012})},\ \Eprint {https://arxiv.org/abs/1202.0806} {arXiv:1202.0806} \BibitemShut {NoStop}%
\bibitem [{\citenamefont {Wolters}\ \emph {et~al.}(2013)\citenamefont {Wolters}, \citenamefont {Sadzak}, \citenamefont {Schell}, \citenamefont {Schr{\"{o}}der},\ and\ \citenamefont {Benson}}]{Wolters2013}%
  \BibitemOpen
  \bibfield  {author} {\bibinfo {author} {\bibfnamefont {J.}~\bibnamefont {Wolters}}, \bibinfo {author} {\bibfnamefont {N.}~\bibnamefont {Sadzak}}, \bibinfo {author} {\bibfnamefont {A.~W.}\ \bibnamefont {Schell}}, \bibinfo {author} {\bibfnamefont {T.}~\bibnamefont {Schr{\"{o}}der}},\ and\ \bibinfo {author} {\bibfnamefont {O.}~\bibnamefont {Benson}},\ }\bibfield  {title} {\enquote {\bibinfo {title} {{Measurement of the ultrafast spectral diffusion of the optical transition of nitrogen vacancy centers in nano-size diamond using correlation interferometry}},}\ }\href {https://doi.org/10.1103/PhysRevLett.110.027401} {\bibfield  {journal} {\bibinfo  {journal} {Physical Review Letters}\ }\textbf {\bibinfo {volume} {110}},\ \bibinfo {pages} {027401} (\bibinfo {year} {2013})}\BibitemShut {NoStop}%
\bibitem [{\citenamefont {Ruf}\ \emph {et~al.}(2019)\citenamefont {Ruf}, \citenamefont {IJspeert}, \citenamefont {van Dam}, \citenamefont {de~Jong}, \citenamefont {van~den Berg}, \citenamefont {Evers},\ and\ \citenamefont {Hanson}}]{Ruf-NanoLett-2019-CoherentNVinMembrane}%
  \BibitemOpen
  \bibfield  {author} {\bibinfo {author} {\bibfnamefont {M.}~\bibnamefont {Ruf}}, \bibinfo {author} {\bibfnamefont {M.}~\bibnamefont {IJspeert}}, \bibinfo {author} {\bibfnamefont {S.}~\bibnamefont {van Dam}}, \bibinfo {author} {\bibfnamefont {N.}~\bibnamefont {de~Jong}}, \bibinfo {author} {\bibfnamefont {H.}~\bibnamefont {van~den Berg}}, \bibinfo {author} {\bibfnamefont {G.}~\bibnamefont {Evers}},\ and\ \bibinfo {author} {\bibfnamefont {R.}~\bibnamefont {Hanson}},\ }\bibfield  {title} {\enquote {\bibinfo {title} {Optically coherent nitrogen-vacancy centers in micrometer-thin etched diamond membranes},}\ }\href {https://doi.org/10.1021/acs.nanolett.9b01316} {\bibfield  {journal} {\bibinfo  {journal} {Nano Letters}\ }\textbf {\bibinfo {volume} {19}},\ \bibinfo {pages} {3987--3992} (\bibinfo {year} {2019})}\BibitemShut {NoStop}%
\bibitem [{\citenamefont {Evans}\ \emph {et~al.}(2016)\citenamefont {Evans}, \citenamefont {Sipahigil}, \citenamefont {Sukachev}, \citenamefont {Zibrov},\ and\ \citenamefont {Lukin}}]{Evans2016}%
  \BibitemOpen
  \bibfield  {author} {\bibinfo {author} {\bibfnamefont {R.~E.}\ \bibnamefont {Evans}}, \bibinfo {author} {\bibfnamefont {A.}~\bibnamefont {Sipahigil}}, \bibinfo {author} {\bibfnamefont {D.~D.}\ \bibnamefont {Sukachev}}, \bibinfo {author} {\bibfnamefont {A.~S.}\ \bibnamefont {Zibrov}},\ and\ \bibinfo {author} {\bibfnamefont {M.~D.}\ \bibnamefont {Lukin}},\ }\bibfield  {title} {\enquote {\bibinfo {title} {{Narrow-Linewidth Homogeneous Optical Emitters in Diamond Nanostructures via Silicon Ion Implantation}},}\ }\href {https://doi.org/10.1103/PhysRevApplied.5.044010} {\bibfield  {journal} {\bibinfo  {journal} {Physical Review Applied}\ }\textbf {\bibinfo {volume} {5}},\ \bibinfo {pages} {044010} (\bibinfo {year} {2016})},\ \Eprint {https://arxiv.org/abs/1512.03820} {arXiv:1512.03820} \BibitemShut {NoStop}%
\bibitem [{\citenamefont {Kasperczyk}\ \emph {et~al.}(2020)\citenamefont {Kasperczyk}, \citenamefont {Zuber}, \citenamefont {Barfuss}, \citenamefont {K\"olbl}, \citenamefont {Yurgens}, \citenamefont {Fl\aa{}gan}, \citenamefont {Jakubczyk}, \citenamefont {Shields}, \citenamefont {Warburton},\ and\ \citenamefont {Maletinsky}}]{Kasperczyk2020}%
  \BibitemOpen
  \bibfield  {author} {\bibinfo {author} {\bibfnamefont {M.}~\bibnamefont {Kasperczyk}}, \bibinfo {author} {\bibfnamefont {J.~A.}\ \bibnamefont {Zuber}}, \bibinfo {author} {\bibfnamefont {A.}~\bibnamefont {Barfuss}}, \bibinfo {author} {\bibfnamefont {J.}~\bibnamefont {K\"olbl}}, \bibinfo {author} {\bibfnamefont {V.}~\bibnamefont {Yurgens}}, \bibinfo {author} {\bibfnamefont {S.}~\bibnamefont {Fl\aa{}gan}}, \bibinfo {author} {\bibfnamefont {T.}~\bibnamefont {Jakubczyk}}, \bibinfo {author} {\bibfnamefont {B.}~\bibnamefont {Shields}}, \bibinfo {author} {\bibfnamefont {R.~J.}\ \bibnamefont {Warburton}},\ and\ \bibinfo {author} {\bibfnamefont {P.}~\bibnamefont {Maletinsky}},\ }\bibfield  {title} {\enquote {\bibinfo {title} {Statistically modeling optical linewidths of nitrogen vacancy centers in microstructures},}\ }\href {https://doi.org/10.1103/PhysRevB.102.075312} {\bibfield  {journal} {\bibinfo  {journal} {Phys. Rev. B}\ }\textbf {\bibinfo {volume} {102}},\ \bibinfo {pages} {075312} (\bibinfo {year}
  {2020})}\BibitemShut {NoStop}%
\bibitem [{\citenamefont {Shevchenko}\ \emph {et~al.}(2024)\citenamefont {Shevchenko}, \citenamefont {Perevislov}, \citenamefont {Nozhkina}, \citenamefont {Oryshchenko},\ and\ \citenamefont {Arlashkin}}]{shevchenko2024high}%
  \BibitemOpen
  \bibfield  {author} {\bibinfo {author} {\bibfnamefont {V.~Y.}\ \bibnamefont {Shevchenko}}, \bibinfo {author} {\bibfnamefont {S.}~\bibnamefont {Perevislov}}, \bibinfo {author} {\bibfnamefont {A.}~\bibnamefont {Nozhkina}}, \bibinfo {author} {\bibfnamefont {A.}~\bibnamefont {Oryshchenko}},\ and\ \bibinfo {author} {\bibfnamefont {I.}~\bibnamefont {Arlashkin}},\ }\bibfield  {title} {\enquote {\bibinfo {title} {High temperature graphitization of diamond during heat treatment in air and in a vacuum},}\ }\href@noop {} {\bibfield  {journal} {\bibinfo  {journal} {Glass Physics and Chemistry}\ }\textbf {\bibinfo {volume} {50}},\ \bibinfo {pages} {69--86} (\bibinfo {year} {2024})}\BibitemShut {NoStop}%
\bibitem [{\citenamefont {Khanaliloo}\ \emph {et~al.}(2015)\citenamefont {Khanaliloo}, \citenamefont {Mitchell}, \citenamefont {Hryciw},\ and\ \citenamefont {Barclay}}]{Khanaliloo2015}%
  \BibitemOpen
  \bibfield  {author} {\bibinfo {author} {\bibfnamefont {B.}~\bibnamefont {Khanaliloo}}, \bibinfo {author} {\bibfnamefont {M.}~\bibnamefont {Mitchell}}, \bibinfo {author} {\bibfnamefont {A.~C.}\ \bibnamefont {Hryciw}},\ and\ \bibinfo {author} {\bibfnamefont {P.~E.}\ \bibnamefont {Barclay}},\ }\bibfield  {title} {\enquote {\bibinfo {title} {{High-Q/V monolithic diamond microdisks fabricated with quasi-isotropic etching.}}}\ }\href {https://doi.org/10.1021/acs.nanolett.5b01346} {\bibfield  {journal} {\bibinfo  {journal} {Nano letters}\ }\textbf {\bibinfo {volume} {15}},\ \bibinfo {pages} {5131--5136} (\bibinfo {year} {2015})}\BibitemShut {NoStop}%
\bibitem [{\citenamefont {Mitchell}\ \emph {et~al.}(2016)\citenamefont {Mitchell}, \citenamefont {Khanaliloo}, \citenamefont {Lake}, \citenamefont {Masuda}, \citenamefont {Hadden},\ and\ \citenamefont {Barclay}}]{Mitchell-DiamondOptomechanicalResonator-Optica-2016}%
  \BibitemOpen
  \bibfield  {author} {\bibinfo {author} {\bibfnamefont {M.}~\bibnamefont {Mitchell}}, \bibinfo {author} {\bibfnamefont {B.}~\bibnamefont {Khanaliloo}}, \bibinfo {author} {\bibfnamefont {D.~P.}\ \bibnamefont {Lake}}, \bibinfo {author} {\bibfnamefont {T.}~\bibnamefont {Masuda}}, \bibinfo {author} {\bibfnamefont {J.~P.}\ \bibnamefont {Hadden}},\ and\ \bibinfo {author} {\bibfnamefont {P.~E.}\ \bibnamefont {Barclay}},\ }\bibfield  {title} {\enquote {\bibinfo {title} {{Single-crystal diamond low-dissipation cavity optomechanics}},}\ }\href {https://doi.org/10.1364/OPTICA.3.000963} {\bibfield  {journal} {\bibinfo  {journal} {Optica}\ }\textbf {\bibinfo {volume} {3}},\ \bibinfo {pages} {963--970} (\bibinfo {year} {2016})}\BibitemShut {NoStop}%
\bibitem [{\citenamefont {Brown}\ \emph {et~al.}(2019)\citenamefont {Brown}, \citenamefont {Chartier}, \citenamefont {Sweet}, \citenamefont {Hopper},\ and\ \citenamefont {Bassett}}]{brown2019cleaning}%
  \BibitemOpen
  \bibfield  {author} {\bibinfo {author} {\bibfnamefont {K.~J.}\ \bibnamefont {Brown}}, \bibinfo {author} {\bibfnamefont {E.}~\bibnamefont {Chartier}}, \bibinfo {author} {\bibfnamefont {E.~M.}\ \bibnamefont {Sweet}}, \bibinfo {author} {\bibfnamefont {D.~A.}\ \bibnamefont {Hopper}},\ and\ \bibinfo {author} {\bibfnamefont {L.~C.}\ \bibnamefont {Bassett}},\ }\bibfield  {title} {\enquote {\bibinfo {title} {Cleaning diamond surfaces using boiling acid treatment in a standard laboratory chemical hood},}\ }\href@noop {} {\bibfield  {journal} {\bibinfo  {journal} {Journal of Chemical Health \& Safety}\ }\textbf {\bibinfo {volume} {26}},\ \bibinfo {pages} {40--44} (\bibinfo {year} {2019})}\BibitemShut {NoStop}%
\bibitem [{\citenamefont {Borselli}, \citenamefont {Johnson},\ and\ \citenamefont {Painter}(2005)}]{ref:borselli2005brs}%
  \BibitemOpen
  \bibfield  {author} {\bibinfo {author} {\bibfnamefont {M.}~\bibnamefont {Borselli}}, \bibinfo {author} {\bibfnamefont {T.~J.}\ \bibnamefont {Johnson}},\ and\ \bibinfo {author} {\bibfnamefont {O.}~\bibnamefont {Painter}},\ }\bibfield  {title} {\enquote {\bibinfo {title} {{Beyond the Rayleigh scattering limit in high-Q silicon microdisks: theory and experiment}},}\ }\href {https://doi.org/10.1364/OPEX.13.001515} {\bibfield  {journal} {\bibinfo  {journal} {Optics Express}\ }\textbf {\bibinfo {volume} {13}},\ \bibinfo {pages} {1515} (\bibinfo {year} {2005})}\BibitemShut {NoStop}%
\bibitem [{\citenamefont {Carmon}, \citenamefont {Yang},\ and\ \citenamefont {Vahala}(2004)}]{ref:carmon2004dtb}%
  \BibitemOpen
  \bibfield  {author} {\bibinfo {author} {\bibfnamefont {T.}~\bibnamefont {Carmon}}, \bibinfo {author} {\bibfnamefont {L.}~\bibnamefont {Yang}},\ and\ \bibinfo {author} {\bibfnamefont {K.~J.}\ \bibnamefont {Vahala}},\ }\bibfield  {title} {\enquote {\bibinfo {title} {Dynamical thermal behavior and thermal self-stability of microcavities},}\ }\href {http://www.opticsexpress.org/abstract.cfm?URI=OPEX-12-20-4742} {\bibfield  {journal} {\bibinfo  {journal} {Opt. Express}\ }\textbf {\bibinfo {volume} {12}},\ \bibinfo {pages} {4742--4750} (\bibinfo {year} {2004})}\BibitemShut {NoStop}%
\bibitem [{\citenamefont {Dychalska}\ \emph {et~al.}(2015)\citenamefont {Dychalska}, \citenamefont {Popielarski}, \citenamefont {Frank{\'o}w}, \citenamefont {Fabisiak}, \citenamefont {Paprocki},\ and\ \citenamefont {Szybowicz}}]{dychalska2015study}%
  \BibitemOpen
  \bibfield  {author} {\bibinfo {author} {\bibfnamefont {A.}~\bibnamefont {Dychalska}}, \bibinfo {author} {\bibfnamefont {P.}~\bibnamefont {Popielarski}}, \bibinfo {author} {\bibfnamefont {W.}~\bibnamefont {Frank{\'o}w}}, \bibinfo {author} {\bibfnamefont {K.}~\bibnamefont {Fabisiak}}, \bibinfo {author} {\bibfnamefont {K.}~\bibnamefont {Paprocki}},\ and\ \bibinfo {author} {\bibfnamefont {M.}~\bibnamefont {Szybowicz}},\ }\bibfield  {title} {\enquote {\bibinfo {title} {Study of cvd diamond layers with amorphous carbon admixture by raman scattering spectroscopy},}\ }\href@noop {} {\bibfield  {journal} {\bibinfo  {journal} {Mater. Sci.-Pol}\ }\textbf {\bibinfo {volume} {33}},\ \bibinfo {pages} {799--805} (\bibinfo {year} {2015})}\BibitemShut {NoStop}%
\bibitem [{\citenamefont {Joe}\ \emph {et~al.}(2024)\citenamefont {Joe}, \citenamefont {Chia}, \citenamefont {Pingault}, \citenamefont {Haas}, \citenamefont {Chalupnik}, \citenamefont {Cornell}, \citenamefont {Kuruma}, \citenamefont {Machielse}, \citenamefont {Sinclair}, \citenamefont {Meesala} \emph {et~al.}}]{joe2024high}%
  \BibitemOpen
  \bibfield  {author} {\bibinfo {author} {\bibfnamefont {G.}~\bibnamefont {Joe}}, \bibinfo {author} {\bibfnamefont {C.}~\bibnamefont {Chia}}, \bibinfo {author} {\bibfnamefont {B.}~\bibnamefont {Pingault}}, \bibinfo {author} {\bibfnamefont {M.}~\bibnamefont {Haas}}, \bibinfo {author} {\bibfnamefont {M.}~\bibnamefont {Chalupnik}}, \bibinfo {author} {\bibfnamefont {E.}~\bibnamefont {Cornell}}, \bibinfo {author} {\bibfnamefont {K.}~\bibnamefont {Kuruma}}, \bibinfo {author} {\bibfnamefont {B.}~\bibnamefont {Machielse}}, \bibinfo {author} {\bibfnamefont {N.}~\bibnamefont {Sinclair}}, \bibinfo {author} {\bibfnamefont {S.}~\bibnamefont {Meesala}}, \emph {et~al.},\ }\bibfield  {title} {\enquote {\bibinfo {title} {High q-factor diamond optomechanical resonators with silicon vacancy centers at millikelvin temperatures},}\ }\href@noop {} {\bibfield  {journal} {\bibinfo  {journal} {Nano Letters}\ }\textbf {\bibinfo {volume} {24}},\ \bibinfo {pages} {6831--6837} (\bibinfo {year} {2024})}\BibitemShut {NoStop}%
\bibitem [{\citenamefont {Mitchell}, \citenamefont {Lake},\ and\ \citenamefont {Barclay}(2019)}]{Mitchell-2019-APLphotonics-DiamondMicrodisks}%
  \BibitemOpen
  \bibfield  {author} {\bibinfo {author} {\bibfnamefont {M.}~\bibnamefont {Mitchell}}, \bibinfo {author} {\bibfnamefont {D.~P.}\ \bibnamefont {Lake}},\ and\ \bibinfo {author} {\bibfnamefont {P.~E.}\ \bibnamefont {Barclay}},\ }\bibfield  {title} {\enquote {\bibinfo {title} {{Realizing Q {\textgreater} 300 000 in diamond microdisks for optomechanics via etch optimization}},}\ }\href {https://doi.org/10.1063/1.5053122} {\bibfield  {journal} {\bibinfo  {journal} {APL Photonics}\ }\textbf {\bibinfo {volume} {4}},\ \bibinfo {pages} {16101} (\bibinfo {year} {2019})}\BibitemShut {NoStop}%
\bibitem [{\citenamefont {Graziosi}\ \emph {et~al.}(2018)\citenamefont {Graziosi}, \citenamefont {Mi}, \citenamefont {Kiss},\ and\ \citenamefont {Quack}}]{graziosi2018single}%
  \BibitemOpen
  \bibfield  {author} {\bibinfo {author} {\bibfnamefont {T.}~\bibnamefont {Graziosi}}, \bibinfo {author} {\bibfnamefont {S.}~\bibnamefont {Mi}}, \bibinfo {author} {\bibfnamefont {M.}~\bibnamefont {Kiss}},\ and\ \bibinfo {author} {\bibfnamefont {N.}~\bibnamefont {Quack}},\ }\bibfield  {title} {\enquote {\bibinfo {title} {Single crystal diamond micro-disk resonators by focused ion beam milling},}\ }\href {https://doi.org/10.1063/1.5051316} {\bibfield  {journal} {\bibinfo  {journal} {Apl Photonics}\ }\textbf {\bibinfo {volume} {3}},\ \bibinfo {pages} {126101} (\bibinfo {year} {2018})}\BibitemShut {NoStop}%
\bibitem [{\citenamefont {Wu}\ \emph {et~al.}(2017)\citenamefont {Wu}, \citenamefont {Sang}, \citenamefont {Teraji}, \citenamefont {Li}, \citenamefont {Wu}, \citenamefont {Imura}, \citenamefont {You}, \citenamefont {Koide},\ and\ \citenamefont {Liao}}]{wu2017reducing}%
  \BibitemOpen
  \bibfield  {author} {\bibinfo {author} {\bibfnamefont {H.}~\bibnamefont {Wu}}, \bibinfo {author} {\bibfnamefont {L.}~\bibnamefont {Sang}}, \bibinfo {author} {\bibfnamefont {T.}~\bibnamefont {Teraji}}, \bibinfo {author} {\bibfnamefont {T.}~\bibnamefont {Li}}, \bibinfo {author} {\bibfnamefont {K.}~\bibnamefont {Wu}}, \bibinfo {author} {\bibfnamefont {M.}~\bibnamefont {Imura}}, \bibinfo {author} {\bibfnamefont {J.}~\bibnamefont {You}}, \bibinfo {author} {\bibfnamefont {Y.}~\bibnamefont {Koide}},\ and\ \bibinfo {author} {\bibfnamefont {M.}~\bibnamefont {Liao}},\ }\bibfield  {title} {\enquote {\bibinfo {title} {Reducing energy dissipation and surface effect of diamond nanoelectromechanical resonators by annealing in oxygen ambient},}\ }\href@noop {} {\bibfield  {journal} {\bibinfo  {journal} {Carbon}\ }\textbf {\bibinfo {volume} {124}},\ \bibinfo {pages} {281--287} (\bibinfo {year} {2017})}\BibitemShut {NoStop}%
\bibitem [{\citenamefont {Metcalf}, \citenamefont {Liu},\ and\ \citenamefont {Abernathy}(2018)}]{metcalf2018improving}%
  \BibitemOpen
  \bibfield  {author} {\bibinfo {author} {\bibfnamefont {T.~H.}\ \bibnamefont {Metcalf}}, \bibinfo {author} {\bibfnamefont {X.}~\bibnamefont {Liu}},\ and\ \bibinfo {author} {\bibfnamefont {M.~R.}\ \bibnamefont {Abernathy}},\ }\bibfield  {title} {\enquote {\bibinfo {title} {Improving the mechanical quality factor of ultra-low-loss silicon resonators},}\ }\href@noop {} {\bibfield  {journal} {\bibinfo  {journal} {Journal of Applied Physics}\ }\textbf {\bibinfo {volume} {123}} (\bibinfo {year} {2018})}\BibitemShut {NoStop}%
\bibitem [{\citenamefont {Tao}\ \emph {et~al.}(2014)\citenamefont {Tao}, \citenamefont {Boss}, \citenamefont {Moores},\ and\ \citenamefont {Degen}}]{Tao2014}%
  \BibitemOpen
  \bibfield  {author} {\bibinfo {author} {\bibfnamefont {Y.}~\bibnamefont {Tao}}, \bibinfo {author} {\bibfnamefont {J.~M.}\ \bibnamefont {Boss}}, \bibinfo {author} {\bibfnamefont {B.~A.}\ \bibnamefont {Moores}},\ and\ \bibinfo {author} {\bibfnamefont {C.~L.}\ \bibnamefont {Degen}},\ }\bibfield  {title} {\enquote {\bibinfo {title} {{Single-crystal diamond nanomechanical resonators with quality factors exceeding one million.}}}\ }\href {https://doi.org/10.1038/ncomms4638} {\bibfield  {journal} {\bibinfo  {journal} {Nature communications}\ }\textbf {\bibinfo {volume} {5}},\ \bibinfo {pages} {3638} (\bibinfo {year} {2014})}\BibitemShut {NoStop}%
\end{thebibliography}

%aipnum4-2.bst 2019-01-14 (MD) hand-edited version of apsrev4-1.bst

\end{document}

\preprint{AIP/123-QED}

\title[]{\textit{Supplementary Material for} Reversing Annealing-Induced Optical Loss in Diamond Microcavities}

% Force line breaks with \\
\author{Vinaya K. Kavatamane}
 \altaffiliation{These authors contributed equally}
\affiliation{Institute for Quantum Science and Technology, University of Calgary, Calgary, Alberta T2N 1N4, Canada}
\author{Natalia C. Carvalho}%
\altaffiliation{These authors contributed equally}
\affiliation{Institute for Quantum Science and Technology, University of Calgary, Calgary, Alberta T2N 1N4, Canada}
\author{Ahmas El-Hamamsy}
 \affiliation{Institute for Quantum Science and Technology, University of Calgary, Calgary, Alberta T2N 1N4, Canada}
 \author{Elham Zohari}
 \affiliation{Institute for Quantum Science and Technology, University of Calgary, Calgary, Alberta T2N 1N4, Canada}
 \affiliation{
Department of Physics, University of Alberta, Edmonton, Alberta T6G 2E1, Canada}
\affiliation{National Research Council Canada, Quantum and Nanotechnology Research Centre, Edmonton, Alberta T6G 2M9, Canada}

\author{Paul E. Barclay}
  \email{pbarclay@ucalgary.ca}
 
 \affiliation{Institute for Quantum Science and Technology, University of Calgary, Calgary, Alberta T2N 1N4, Canada}


\maketitle

%\onecolumn

\section{High-temperature and high-vacuum annealing}

The annealing experiments were carried out in a home-built vacuum annealing setup consisting of a vacuum chamber that houses a heater (Boralectric heater HTR1002, tectra GmBH). The heater is capable of reaching up to 1200$^\circ$C and is controlled by an HC3500 power supply with a PID temperature controller. The sample was placed in a custom-made molybdenum holder that sits on top of the heating element to ensure proper and uniform heating of the sample while also securing the sample from mechanical vibrations due to the turbo pump. A heat shield made of molybdenum is placed on top of the heater to avoid the excessive heating of the chamber due to radiation.
The vacuum chamber is a four-way cross (C-0800, Kurt J. Lesker, USA) with its ports used as follows: P1: turbo pump (Pfeiffer), P2: ion pump (Varian, Star Cell), P3: used for attaching the heater, and P4: used for loading (unloading) the sample into (from) the chamber. P4 is also used to measure the chamber pressure using a pressure gauge via a T-shaped vacuum flange (see Fig. 1 (b) in the main text). The T-flange's other port is connected to a low-pressure N$_\text{2}$ line aided by an angle valve for venting the chamber, while the T-flange's last port is blocked with a blank flange.\\ 

Before each annealing, the chamber is brought to a base vacuum level of 10$^{-9}$ mbar by a combination of the following methods: pumping the chamber using a turbo pump, baking the whole setup around 100-150$^\circ$C using external heaters, and a final pumping using the ion pump. For annealing experiments, the heater temperature was slowly ramped from room temperature, as explained in the main article, while continuously monitoring the chamber pressure.

\section{Surface treatment}

Our surface treatment of the diamond sample involves the standard method of boiling it in an equal mixture of three acids, H$_2$SO$_4$, HNO$_3$, and HClO$_4$\cite{brown2019cleaning}. A bath of the aforementioned tri-acid mixture containing the sample was subjected to a constant temperature of 250$^\text{o}$ C in a glass flask fitted with a water-cooled reflux condenser in a dedicated perchloric acid-safe fume hood. Each boiling was maintained for about 120 minutes, followed by sample extraction. The sample was then repeatedly rinsed in ultrapure water (ASTM type II grade), and finally dried by gently blowing N$_2$ gas to remove any water residues.

\section{Optical cavities}

\subsection{Quality factor}

\begin{figure*}[h!] 
\centering\includegraphics[width=\linewidth]{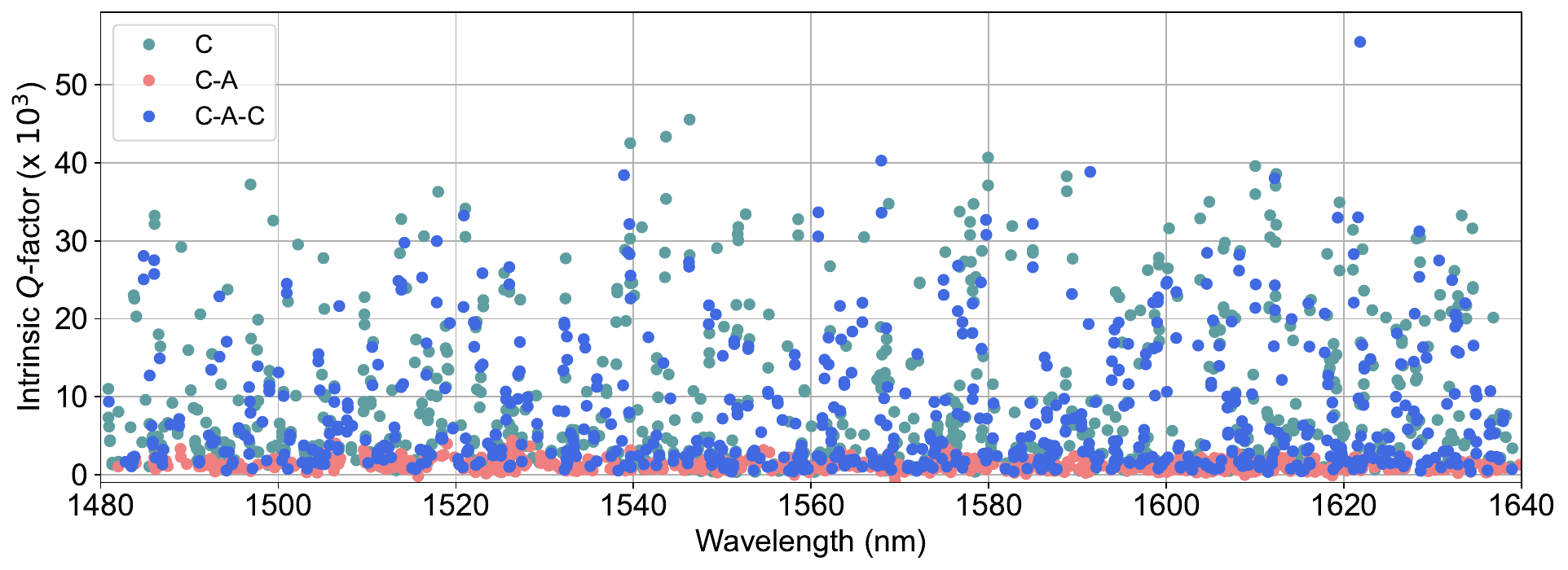}
\caption{Scatter plot of intrinsic quality factor of all measured modes of microdisks ranging from 5 to 8 $\mu$m at different stages of processing: clean pre-annealing (C), annealed (C-A) and clean post-annealing (C-A-C)). Multiple microdisks of the same size were measured.}
\label{s1}
\end{figure*}

\begin{figure*}[h!] 
\centering\includegraphics[width=\linewidth]{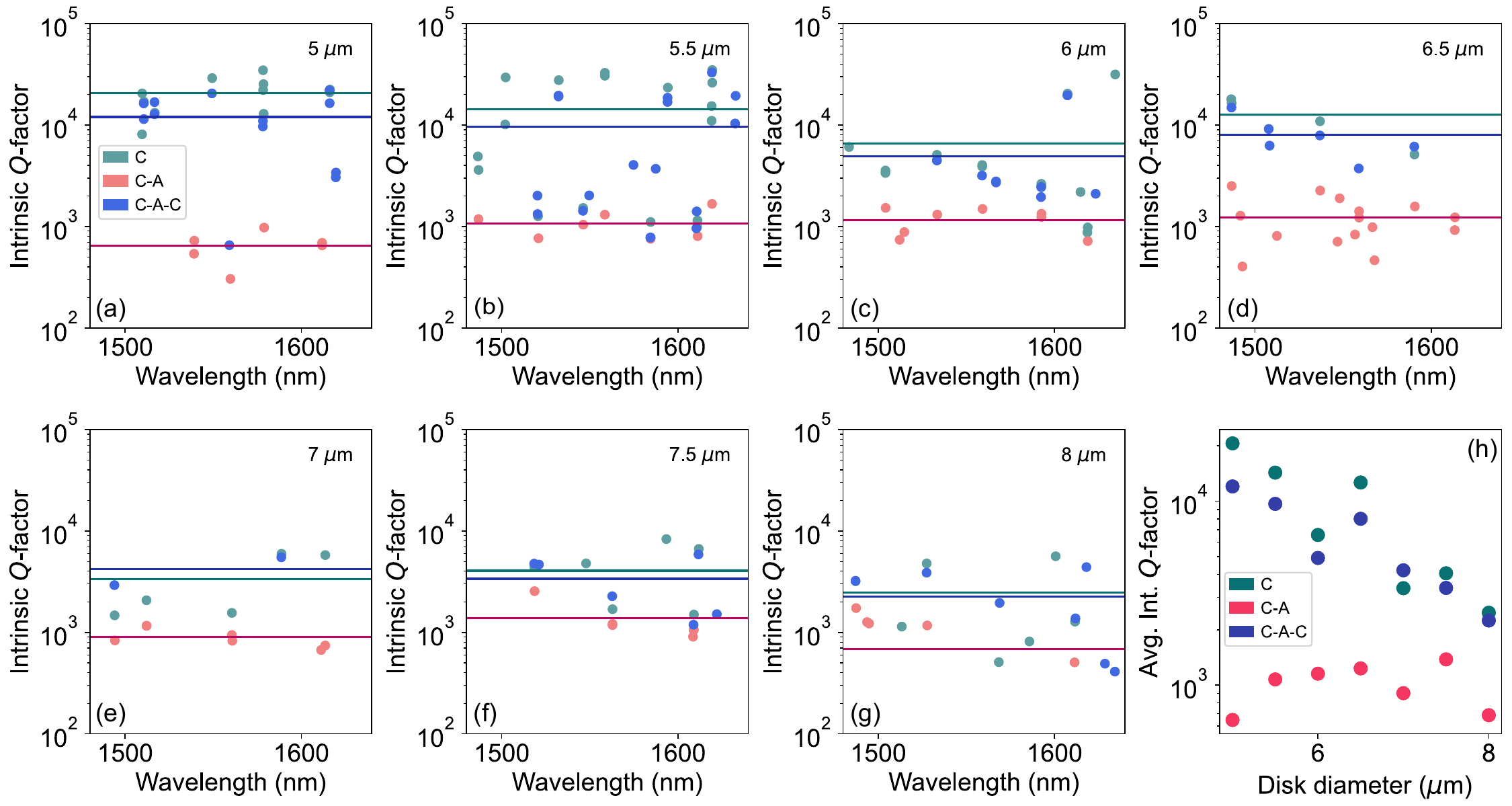}
\caption{(a) - (g): Intrinsic quality factor of modes measured in microdisks of different diameters as indicated in the top right of each plot. Solid lines show the average of the modes measured in the three conditions: clean pre-annealing (C), annealed (C-A) and clean post-
annealing (C-A-C). (h) Average intrinsic quality factors as a function of the microdisk diameter.  }
\label{s2}
\end{figure*}

To investigate whether high-temperature high-vacuum annealing and tri-acid cleaning affect optical loss of the diamond optical cavities, the spectroscopic studies of modes of multiple microdisks with diameters ranging from 5 to 8\,$\mu$m were recorded. The extracted intrinsic optical quality factors, $Q_\text{opt}^i$, of several modes of each of the measured microdisks are presented in Fig.\ \ref{s1}. In Figs.\ \ref{s2} (a) - (g), we show $Q_\text{opt}^i$ of modes from individual microdisks of different diameters (data points) superposed with their average. In Fig.\ \ref{s2} (h), we present the average $Q_\text{opt}^i$ as a function of microdisk diameter, as presented in Fig.\ 3 (a) in the main text. Note that in Fig. 3 (a) and Fig.\ \ref{s1}, data from multiple microdisks of the same diameter are included while Fig.\ \ref{s2} contains only one microdisk of each size.

In all of our analysis, we use the intrinsic quality factor as it removes the influence of coupling to the fiber taper waveguide. Intrinsic quality factor is determined from $1/Q_\text{opt}^{i} = 1/Q_\text{opt} - 1/Q_\text{opt}^{e} - 1/Q_\text{opt}^{p}$, where $1/Q_\text{opt}^{e}$ and $1/Q_\text{opt}^{p}$ are proportional to coupling rates of the mode to the transmission channel of the fiber taper, and to parasitic loss channels induced by the fiber taper, respectively. From the resonance contrast and $1/Q_\text{opt}$ obtained from the fits, we can determine $1/Q_\text{opt}^{e}$, and in turn, $1/Q_\text{opt}^{i} + 1/Q_\text{opt}^{p}$. We next assume that for a singlet resonance the fiber taper to microdisk coupling is ideal, so that $1/Q_\text{opt}^{p} = 0$, while for a doublet resonance, the only parasitic loss is into the backward propagating mode of the fiber taper, so that $1/Q_\text{opt}^{p} = 1/Q_\text{opt}^e$. 

\subsection{Frequency shift}

\begin{figure*}[h!] 
\centering\includegraphics[width=\linewidth]{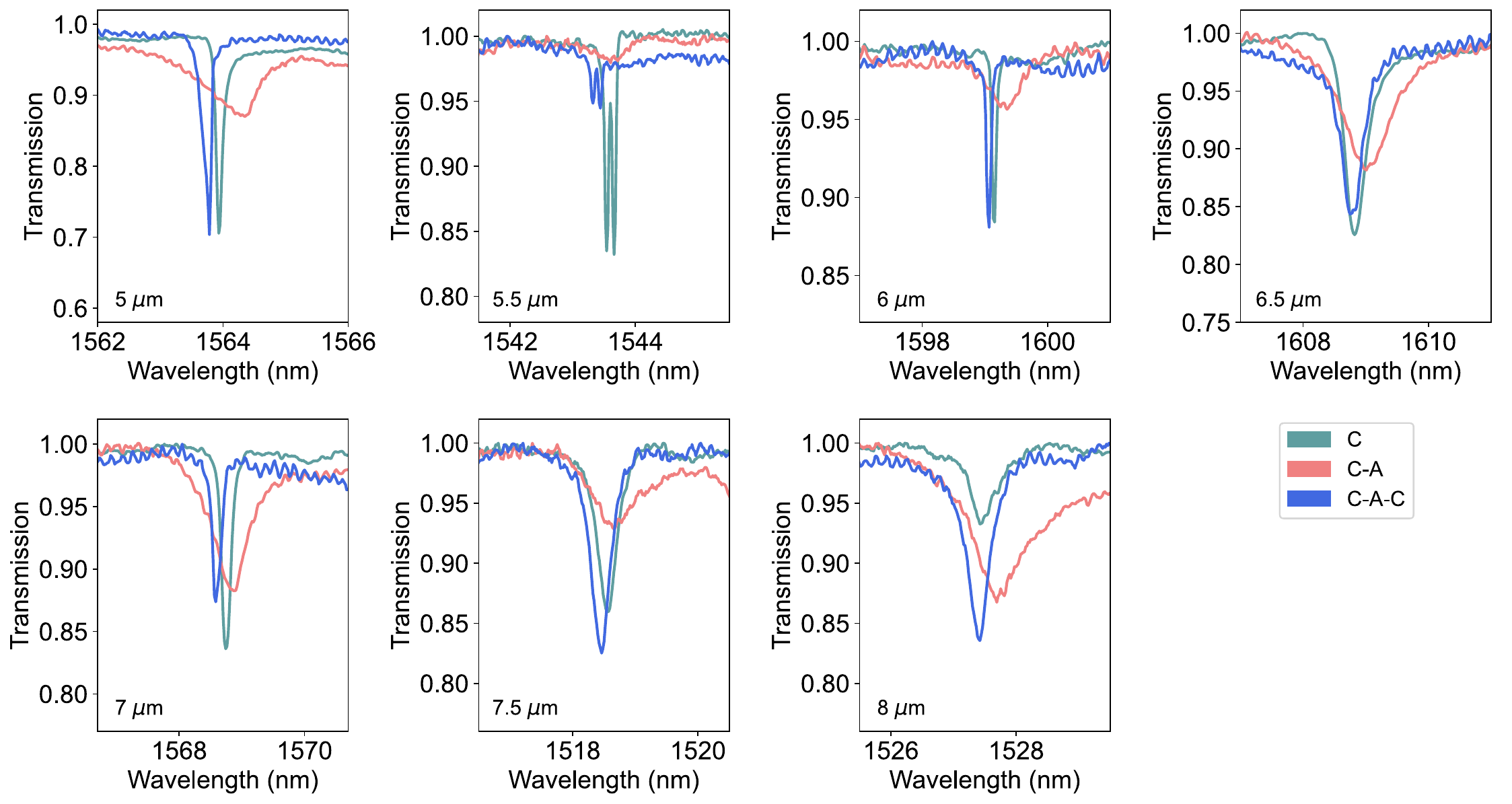}
\caption{Typical normalized fiber taper transmission spectra when probing modes from different diameter microdisks. The diameters of these microdisks are indicated in the lower left of each plot. The resonances were measured in the three processing conditions: clean pre-annealing (C), annealed (C-A) and clean post-annealing (C-A-C).}
\label{s3}
\end{figure*}

The measured optical resonances showed slight red and blue shifts after annealing and cleaning, respectively. This shifting at the sub-nanometer scale is consistent with the creation and removal of a thin layer of material during the procedures. However, it is noticeable that the final (C-A-C) resonance wavelength is on the blue side of the pre-annealed sample (C), indicating that a small portion of pristine diamond was etched. Fig.\ \ref{s3} shows typical resonance shifts for each microdisk diameter.     

\subsection{Modeling photothermal effects and determining optical absorption rates}

\begin{figure} 
\centering\includegraphics[width=0.8\linewidth]{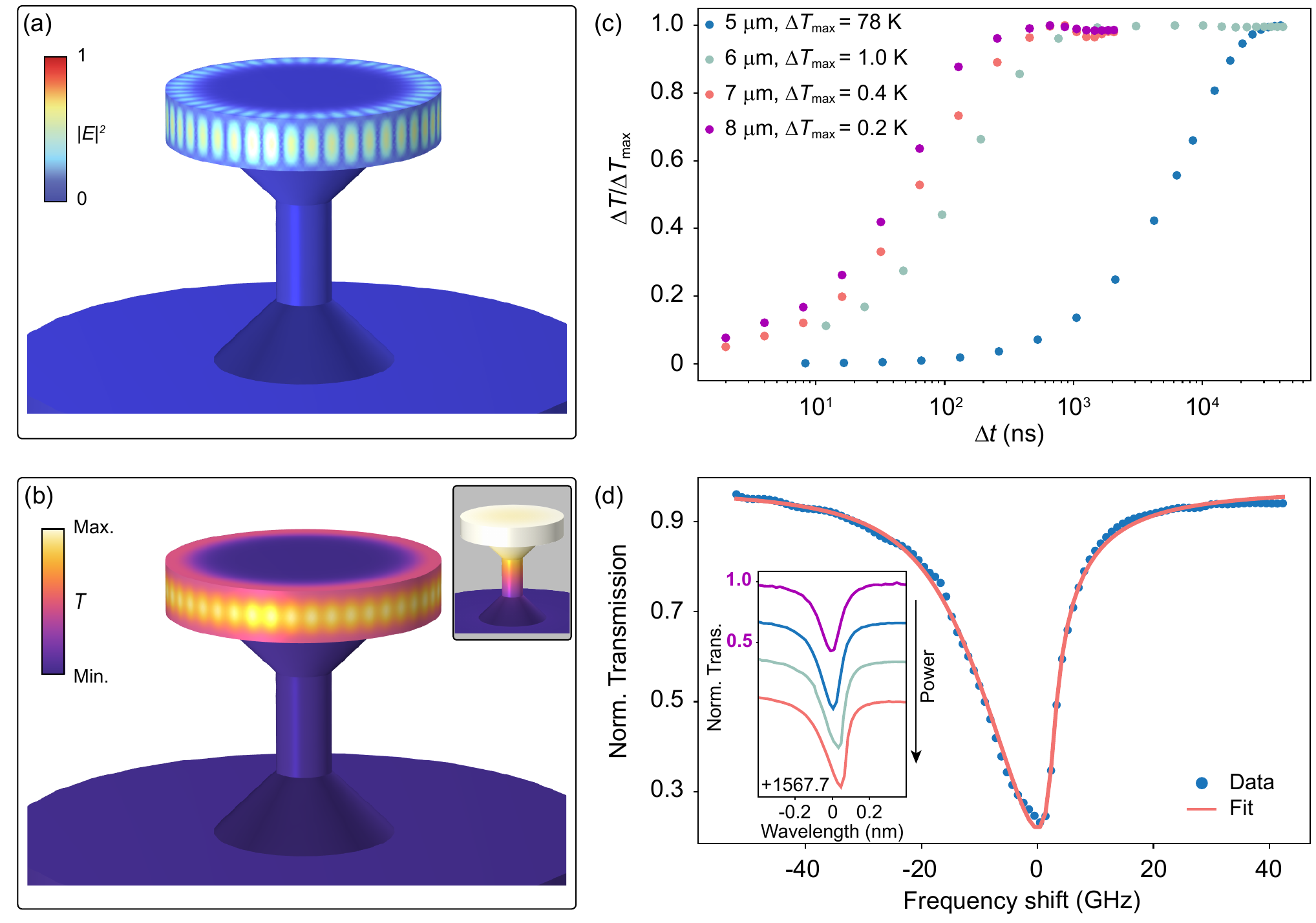}
\caption{Finite element method simulation (COMSOL) of the optical cavity: (a) optical mode and (b) temperature immediately after the optical power is turned on (microdisk of 6\,$\mu$m diameter). The inset shows the temperature distribution after 40\,$\mu$s. (c) The simulation results show the microdisk's temperature increase, $\Delta T$, as a function of time, $\Delta t$, where $\Delta T_{\text{max}}$ is the maximum temperature variation for each disk. The data for microdisks of intermediate diameters are omitted for simplicity. (d) Typical modified Lorentzian due to optical thermal absorption and model fit. Inset: resonance distortion with increasing optical power. } 
\label{s4}

\end{figure}

Despite diamond's large thermal conductivity, the high circulating power enabled by the high $Q_\text{opt}$ of the microdisk modes studied here, combined with their small mode volume and corresponding high per-photon field intensity, allows small amounts of optical absorption to lead to heating of the microcavity. This heating results in thermal expansion and a change in refractive index due to the thermo-optic effect \cite{Mitchell-DiamondOptomechanicalResonator-Optica-2016}, which can lead to a red-shift of the resonance frequency of the microdisk modes. If this intracavity power-dependent shift is non-negligible compared to the mode linewidth, it manifests in a distortion of the nominally Lorentzian lineshape \cite{ref:carmon2004dtb}.  If the thermo-optic properties of the cavity mode can be measured or estimated, this shift can be used extract information regarding the magnitude of optical absorption\cite{Wang2021using}.

To model a transmission profile subject to photothermal effects, we assume that they are instantaneous on the timescale of the measurements performed here, and introduce a resonance shift, $\Delta \omega$ to the Lorentzian function describing the normalized cavity response:

\begin{equation}
\mathcal{L}_{mod}(\omega) = \frac{(\gamma_T/2)^2}{(\omega -\omega_0 + \Delta \omega)^2 +(\gamma_T/2)^2},
\end{equation}
where,
\begin{equation}
\begin{split}
\frac{\Delta \omega}{\omega_0} =  \eta \frac{\gamma_{abs}}{\gamma_i} P_i (1 - T),\\
\eta = \frac{1}{n} \frac{dn}{dT} \frac{1}{\kappa_{th}}.
\end{split}
\label{e2}
\end{equation}

Here,  $\gamma_T = \gamma_i + \gamma_e$ is the total linewidth of the optical mode, $\gamma_i$ ($\gamma_e$) is the intrinsic (extrinsic) component, and $\gamma_{abs}$ is the optical absorption rate. The power of the laser at frequency $\omega$ input to the fiber taper waveguide is $P_i$, the normalized cavity transmission is $T$, and the resonance frequency is $\omega_0$. The second part of Eq.\ \ref{e2} is composed by the
  
diamond's thermo-optic coefficient $\frac{1}{n} \frac{dn}{dT}$ \cite{ruf2000temperature} and the thermal conductance between the microdisk cavity and the diamond substrate, $\kappa_{th}$. The latter quantity was estimated numerically using a finite element model (COMSOL) to calculate the resonator equilibrium temperature ($T$) at a given absorbed  power, $P_{abs}$ (see Fig.\ \ref{s4} (a) and (b)). To obtain $\kappa_{th}$ (= $\Delta T/P_{abs}$) 
 
for different microdisk diameters, it was necessary to estimate the pedestal geometries using the quasi-isotropic etching rates given by Khanaliloo \textit{et al} \cite{Khanaliloo2015}. The simulated change in temperature for 0.5 mW absorbed power is shown in Fig.\ \ref{s4} (c). 

In this work, the thermal expansion contribution to the temperature-dependent optical dispersion is smaller than the thermo-optic effect and is not accounted for in the model. Its effect could slightly reduce the estimated $\gamma_{abs}$. 

The fit of this model to a typical cavity mode is shown in Fig.\ \ref{s4} (d). The inset shows the resonance lineshape for increasing optical input power, evidencing the photothermal effects as more optical power is absorbed. The data was fit with the normalized transmission given by: 
\begin{equation}
T = (1 - T_0)\mathcal{L}_{mod} = \frac{4 c_0}{(c_0 + 1)^2} \mathcal{L}_{mod},
\end{equation}
where $T_0$ is the on-resonance transmission given by $T_0 = \left((1 - \gamma_e/\gamma_i)/(1+\gamma_e/\gamma_i)\right)^2$ (for $\gamma_e < \gamma_i$), $c_0 = \gamma_e/\gamma_i$, and $\gamma_{abs}$, $\gamma_{i}$ and $\gamma_{e}$ are free fitting parameters. Subsequently, we assumed $\gamma_{i} = \gamma_{sca} + \gamma_{abs}$, where $\gamma_{sca}$ is related to the radiation losses due to surface scattering. Our results demonstrate that $\gamma_{sca} \gg \gamma_{abs}$ for the measured microdisks, revealing that surface roughness is the dominant source of optical loss. 

\section{Raman spectroscopy setup}

\begin{figure}[h!] 
\centering\includegraphics[width=0.5\linewidth]{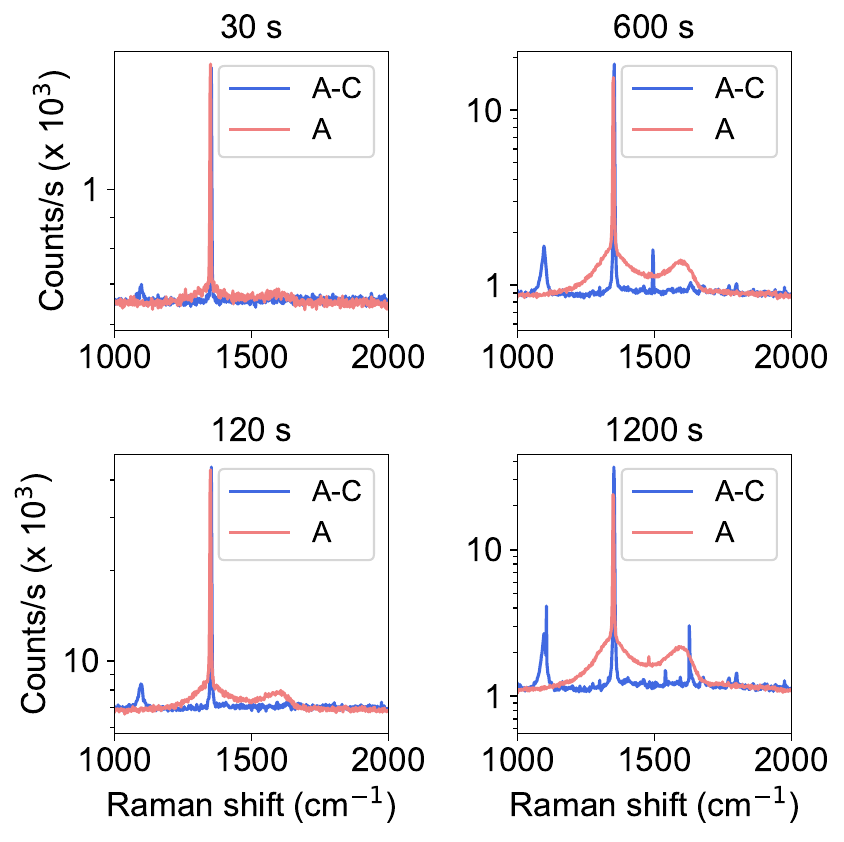}
\caption{Raman spectroscopic data on the annealed sample (A) and cleaned sample post-annealing (A-C) for various signal acquisition intervals. Sharp spikes, besides the native diamond signal, in A-C for higher time intervals are due to noise.}
\label{s5}

\end{figure}

To identify the presence of an amorphous layer post-annealing, we first attempted both standard (wide-angle) and grazing incidence X-ray diffraction spectroscopy on the annealed diamond surface, but found that these methods were insensitive to finding any amorphous signal. \\
To better understand the roughness introduced by the annealing process, we then performed Raman spectroscopic studies on the diamond sample at various stages of the study. This non-destructive technique is based on photon scattering and is used to characterize the surface chemistry of the sample surfaces.  The home-built setup consisted of an excitation laser with a wavelength of around 635\,nm which was cleaned up using two narrow band pass filters (Semrock, FF01-637/7-25) and focused on the sample using an objective lens (Olympus LCPLN50XIR, NA 0.65). The resulting Raman signal from the sample was collected using the same objective and separated from the excitation beam using a combination of 50:50 beam splitter and two 650\,nm longpass filters (Semrock, BLP01-635R-25). The signal was then coupled to a single mode optical fiber (SMF, black jacketed to avoid stray light entry) with a mode field diameter of $\sim$4 $\mu$m at $\sim$50 $\%$ coupling efficiency using an aspheric lens (Thorlabs, C220TMD-B). The signal was then routed to a spectrometer (Princeton Instruments, SP2750i) equipped with an air-cooled detector (CCD). The spectrometer was operated at a resolution was 300 gratings/mm that results in a spectral resolution was $\sim$1.7\,cm$^{\text{-1}}$. Figure \ \ref{s5} shows the Raman spectra of annealed sample (A) and freshly cleaned (A-C) sample post-annealing for different signal acquisition intervals. A confocal microscope arrangement was used to ensure the collected Raman signal was coming from the surface of the diamond. To this end, the same microscope objective attached to a 3-axis nanopositioning piezo-stage scans the sample surface
 while the reflected light from the sample (without 650\,nm filters) was directly used to image (in a 2D raster scan) the sample surface in the confocal mode. Here, the 
SMF acts like a spatial filter and helps select the right measurement depth for collecting the Raman spectra. Even though the axial (depth) resolution of the microscope is lower than the thickness of the non-diamond layer formed due to annealing, the presence (absence) of non-diamond peak post-annealing (pre-annealing) confirms that the Raman method is sensitive to surface composition.